# D2D: Digital Archive to MPEG-21 DIDL

Suchitra Manepalli, Giridhar Manepalli, Michael L. Nelson

Old Dominion University, Department of Computer Science, Norfolk VA 23508 USA
{smanepal, gmanepal, mln}@cs.odu.edu

**Abstract.** Digital Archive to MPEG-21 DIDL (D2D) analyzes the contents of the digital archive and produces an MPEG-21 Digital Item Declaration Language (DIDL) encapsulating the analysis results. DIDL is an extensible XML-based language that aggregates resources and the metadata. We provide a brief report on several analysis techniques applied on the digital archive by the D2D and provide an evaluation of its run-time performance.

## 1 INTRODUCTION

With the increasing trend for digitizing information, technologies are invented to support the growth of digital information. Technologies were adopted, modified and relinquished in a short span, thereby resulting in a huge loss of digital information. The goal of digital preservation is to enable access mechanisms over digital information, by making it transparent to changing technologies over a period of time. Various analysis techniques on digital content exploring its context, provenance and reference are necessary to understand and effectively preserve it. Digital Archive to MPEG-21 DIDL (D2D) is a tool for applying various analysis techniques on a digital archive and producing a Digital Item Declaration Language (DIDL). MPEG-21 is an open standards-based multimedia framework designed to support interoperability of content across communities [8]. MPEG-21 DIDL is XML based language designed to define digital items. DIDL schema is extensible and allows the integration of metadata with the digital items. The tool analyzes the different digital items present inside the archive by applying various techniques on the bit stream of each digital item. The metadata produced by these analysis techniques are stored inside the DIDL. DIDL therefore stands as a single source for future reference to the digital items in the archive, and the analysis reports. The DIDL produced conforms to a well-defined schema, thus presenting itself as an excellent resource for extending analysis using automated programs.

The work presented here builds on past work. Using MPEG-21 DIDL in a digital library context was first explored by the Los Alamos National Laboratory [3]. The use of MPEG-21 DIDL in an aggregative preservation model was first explored in the Archive Ingest and Handling Test [9].

## 2 DIGITAL PRESERVATION

A digital archive, in the present context, is a file system based collection of files (digital items). The analysis captures the context, provenance, and additional metadata pertaining to the bit stream of the archived digital items [4]. Using the Open Archival Information System (OAIS) terminology, this metadata is termed as PDI (Preservation Description Information) [1]. Although multiple techniques may be applied to the digital item, the tool only implemented few of them. Nevertheless, D2D is extensible to any other techniques as long as it is conformant to the provided API. The metadata resulted from each analysis is classified into four entities of PDI namely Provenance, Context, Reference and Fixity [5]. The classified information is then pushed into the PDI of the corresponding digital item. The digital item as well as its PDI is then integrated into DIDL.

PDI is preservation description information, necessary to preserve the content of the digital item. It captures the past and present state of digital item. The four entities in PDI are Provenance, Context, Reference and Fixity. Provenance is the history of a digital item. In other words, it defines when a digital item was created, when it was last modified, who the creator is etc. This information is used to maintain the integrity of the digital item's past for reasons of public accessibility. Context information describes the scope and environment of the digital item. It explains when and how the digital object was created etc. Reference describes the identification systems used for the digital item. A typical reference includes assigned identifiers for the digital item. Fixity provides necessary information for validating and maintaining the integrity of the digital content. Fixity ensures the content information from unauthorized alterations.

**Table 1: Examples of PDI Types (from [1]).**

| Content Information Type | Provenance | Context | Reference | Fixity |
|---|---|---|---|---|
| Digital Library Collections | Pointer to original document, Pointer to earlier versions of the document | Pointers to related documents in original environment at the time of publication | Persistent identifiers | Digital Signature, Checksum |

### 2.1 Analysis Techniques

Applying various analysis techniques on the digital items in the digital archive produces metadata pertaining to the digital items. The metadata thus produced, is the primary source for understanding the content information (beside the content itself). For each analysis, the following digital item sample is considered.

**Table 2: Sample.txt.**

| |
|---|
| I am a man of few words, and account the time for action. But, here we go the little saga for this episode as it unravels.<br><br>I engaged myself in the profound sciences, mathematics, to educate myself and to enrich my mind with "thoughts that matter". |

**MD5 Analysis**

MD5 is a cryptographic hash algorithm that produces a 128-bit hash value. The algorithm generates the hash value by dividing the bit stream of the content into multiple blocks of 512-bits. Each block of 512-bits modify the state of four 32-bit variables. The final state of these four variables constitutes the hash value [7]. The idempotency of the message digest helps maintain the integrity of the original data object content from unauthorized alterations.

For the analysis, UNIX command line utility "md5sum" is used to compute the message digest for a given bit stream.

**Table 3: Output of MD5Sum analysis in hexadecimal system.**

| |
|---|
| a5e69b225922a3a6cd46a20c94ecf1c1 Sample.txt |

**Mapping**

As mentioned above, message digests are used in verifying the integrity of the digital item. The message digests is a time invariant for any given digital item, thus it is categorized as 'fixity'.

**Table 4: Mapping MD5 analysis to PDI Entities.**

| Analysis | PDI Entity | Output |
|---|---|---|
| MD5 | Fixity | a5e69b225922a3a6cd46a2c94ecf1c1 |

**File Analysis**

Any digital item existing on a file system may be classified based on three broad features: the purpose of the file, the type of the file, and the format of the file. Purpose could be a content storage block, a special connection to a remote entity, a link to different entities on the same file system etc. Type could be a machine-language file, a specific interpreter input etc. Format is the file format or the language. Any information in this regard pertaining to each digital item is important for the preservation of the item.

For the specific analysis purpose, the 'file' command available on UNIX machines is used. This command performs three different tests: file system tests, magic number tests, and language tests [2]. File System tests focus on the purpose of the file, magic number tests focus on the type and language tests try to understand the character set and the language of the file.

**Table 5: Output of file analysis.**

| |
|---|
| text/plain; charset=us-ascii<br>ASCII English text |

**Mapping**

The file analysis is based on the current file system, the definitions of encoding formats, and the existence of operation systems. Since the analysis report is dependent on external factors that may change over a period of time, the result is appropriately categorized as a 'context' of the digital item. The table below shows the mapping:

**Table 6: Mapping file analysis to PDI Entities.**

| Analysis | PDI Entity | Output |
|---|---|---|
| File | Context | text/plain; charset=us-ascii |

**Raw Characters Analyzer**

A digital item consists of a bit stream.Some of the encoding formats available for the digital item are ASCII (7-bit), UTF-8, UTF-16 and in general octet-stream. Most of the available terminals support ASCII encoding and is usually defined as printable. The values that fall into octet-stream are specific application dependent. If the application that defined the range of octet values taken by the digital item is obsolete, the digital item itself cannot be comprehensible. For the aforementioned reason, it is preferable to identify the printable range from a digital item as it is non-dependent on any application.

The 'strings' command looks for any printable characters from the specified file, regardless of the file format in normal files and looks at the initialized data space of object files.

**Table 7: Output of strings analysis.**

| |
|---|
| am a man of few words, and account the time for action. But, here we go the little saga for this episode as it unravels.<br>I engaged myself in the profound sciences, mathematics, to educate myself and to enrich my mind with "thoughts that matter". |

**Mapping**

The bytes with the printable range may have relevance in two different directions. One, the content in the digital item itself may emphasize it textually. Two, the digital item may be an octet-stream specific to an application e.g. PDF stream. The printable characters define the syntax guidelines of the format and may prove to be useful in identifying the digital format. Considering the above two reasons, the printable characters report is categorized into 'fixity'.

**Table 8: Mapping raw characters analysis to PDI Entities.**

| Analysis | PDI Entity | Output |
| --- | --- | --- |
| Strings | Fixity | am a man of few words, and account the time for action. But, here we go the little saga for this episode as it unravels.<br>I engaged myself in the profound sciences, mathematics, to educate myself and to enrich my mind with "thoughts that matter". |

**JHOVE Analysis**

The JSTOR/Harvard Object Validation Environment (JHOVE) provides functions to perform format specific identification, validation, and characterization of digital objects [6]. Whereas the "file" command can provide limited knowledge about many formats, JHOVE provides greatly detailed information about a smaller number of formats. Using both "file" and JHOVE provides both breadth and depth about file introspection. Format identification determines the format of the digital item. Format validation determines the level of conformance by the digital item with the standards of the format. In a particular case of XML format, format validation verifies the validation of digital item at two different levels, well-formity and validity. Format characterization determines the significant characteristics of the digital objects like the checksum, size, format, last modified etc.Metadata produced using JHOVE is categorized into the four entities of PDI. The following tables illustrate the JHOVE output and the categorized PDI for an example text file.

**Table 9: Output of JHOVE analysis.**

```
<?xml version="1.0" encoding="UTF-8"?>
<jhove xmlns:xsi="http://www.w3.org/2001/XMLSchema-instance"
xmlns="http://hul.harvard.edu/ois/xml/ns/jhove"
xsi:schemaLocation="http://hul.harvard.edu/ois/xml/ns/jhove
http://hul.harvard.edu/ois/xml/xsd/jhove/jhove.xsd" name="Jhove" release="1.0 (beta
2)" date="2004-07-19">
 <date>2005-12-01T13:19:41-05:00</date>
 <repInfo uri="test.txt">
  <reportingModule release="1.0" date="2004-05-05">ASCII-hul</reportingModule>
  <lastModified>2005-12-01T13:21:49-05:00</lastModified>
  <size>250</size>
  <format>ASCII</format>
  <status>Well-formed and valid</status>
  <mimeType>text/plain; charset=US-ASCII</mimeType>
  <properties>
   <property>
    <name>ASCIIMetadata</name>
    <values arity="List" type="Property">
    <property>
     <name>LineEndings</name>
     <values arity="List" type="String">
      <value>LF</value>
     </values>
    </property>
    </values>
```

```
   </property>
  </properties>
  <checksums>
   <checksum type="CRC32">d0d525af</checksum>
   <checksum type="MD5">a5e69b225922a3a6cd46a2c94ecf1c1</checksum>
   <checksum
type="SHA-1">91cad9d8c2336af58ae04c2876cb9ad61de5510</checksum>
  </checksums>
 </repInfo>
</jhove>
```

**Mapping**

As mentioned above, JHOVE produces metadata pertaining to identification, validation and characterization. The output of JHOVE therefore corresponds to multiple entities inside the PDI. For example, checksum produced by different algorithms namely CRC32, SHA-1, and MD5 can be classified as 'fixity' for aforementioned reasons. Mime-type and format are classified as 'context', based on explanation given in file analysis. From the definition of 'reference', it is well understood that URI falls into this category. Last-Modified Date is defined as the 'provenance' of the digital item. The following table summarizes the mapping of JHOVE analysis to PDI entities.

**Table 10: Mapping JHOVE analysis to PDI Entities.**

| Analysis | PDI Entity | Output |
|---|---|---|
| Last Modified | Provenance | 2005-12-01T13:21:49-05:00 |
| Mime-type | Context | text/plain; charset=US-ASCII |
| Format | Context | ASCII |
| Status | Context | Well-formed and valid |
| URI | Reference | test.txt |
| CRC32 | Fixity | d0d525af |
| MD5 | Fixity | a5e69b225922a3a6cd46a2c94ecf1c1 |
| SHA-1 | Fixity | 91cad9d8c2336af58ae04c2876cb9ad61de5510 |
| Size | Fixity | 250 |

**Fred**

Format Registry Demonstration demonstrates a simple format registry service. Fred is proof of concept prototype for GDFR. Fred provides information regarding registry formats, the characterization of formats for analysis purposes. [2]

**Table 11: Output of FRED analysis.**

```
info:gdfr/fred/f/ascii
```

**Mapping**

FRED analysis results in a URI of the corresponding entry in FRED. Since, the URI is associated with the format of the digital item; it is categorized as 'context'. The following table illustrates the mapping.

**Table 12: Mapping FRED analysis to PDI Entities.**

| Analysis | PDI Entity | Output |
|---|---|---|
| FRED | Context | info:gdfr/fred/f/ascii |

**Java's security API**

Java's Message Digest class provides different algorithms to calculate the checksums of digital objects content. The two most popular algorithms provided by Java are MD5 and SHA-1. The two checksums provides the basis to maintain the integrity of data. In addition to the MD5 command on UNIX or Linux, these algorithms from Java provide an additional level of confidence.

The metadata produced by these analysis techniques is categorized into the four entities of PDI. The PDI of each digital item is used in the second phase for preservation analysis

**Table 13: Output of Java's Message Digest analysis.**

| |
|---|
| SHA-1 91CAD9D8C2336AF58AE04C2876CB9AD61D0E5510<br>MD5 A5E69B225922A3A6CD46A20C94ECF1C1 |

## 3 MPEG-21 DIDL

MPEG-21 is an open standards-based multimedia framework designed to support interoperability of content across communities [8]. The content in the present context is the entity or 'work' that is being managed, described, exchanged, and collected. The content along with some additional relevant information is hereby referred as Digital Item. MPEG-21 Digital Item Declaration is a data model used to describe the set of abstract terms, or concepts to define the digital item.

MPEG-21 has an abstract data model that does not have specific semantics encoded in its declaration language. There are many nuances, but the most important concept is the definition of Containers, Items, Components and Resources. As Figure 1 shows, a DIDL contains at least one container and containers can be recursively defined. Containers eventually hold one or more Items, and Items hold one or more Components or Items. Components hold one or more Resources. Resources are the leaf nodes in the data model; they either contain URIs (by-reference representation) for data objects (PDFs, MPEGs, HTML, etc.) or contain the actual data objects in base64 encoded XML (by-value representation). Resources are the ultimate "thing" that we wish to convey, and the additional infrastructure allows the expression of the hierarchy and relationships between multiple data objects. Although a Component can contain multiple Resources, by definition the Components are considered to be equivalent representations; multiple Resources are generally specified in order to have by-reference or by-value representations, or possibly different encodings (e.g., .zip vs. .gz) of the same data object.

The other important consideration for understanding MPEG-21 DIDL is that every level in the hierarchy except for Resources can have extensible Descriptor elements (multiple Resources are bound together in a single Component, and the Component's Descriptors apply equally across all the Resources in the Component). Descriptors are simply wrapper elements; they can contain any XML encoded data. Some of the standard Descriptors that are defined by MPEG-21 include digital item identifiers (DII), digital item processing (DIP), rights expression language (REL), digital item relations (DIR), and digital item creation date (DIDT).

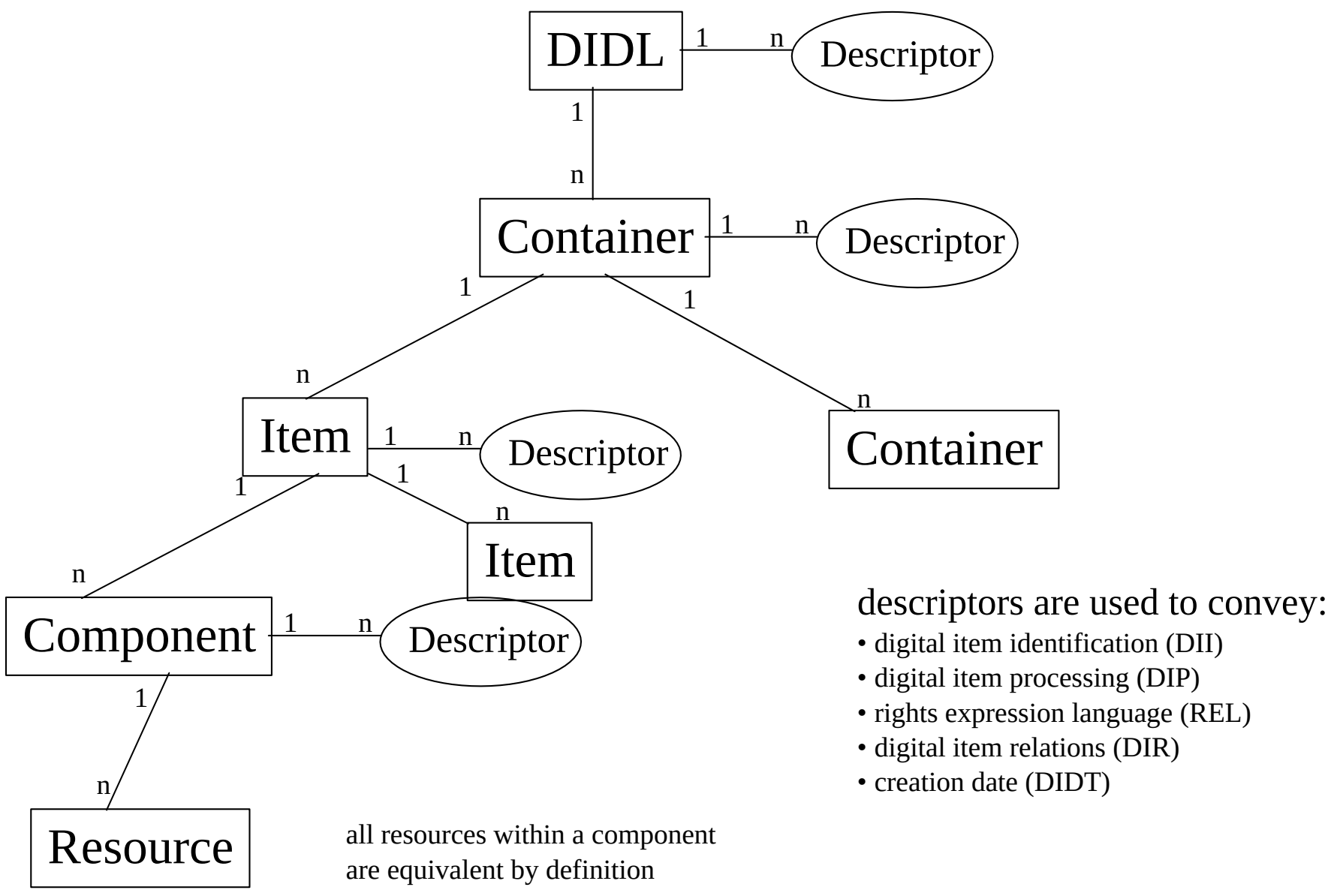

**Figure 1. Highlights of the MPEG-21 Data Model.**

## 4 D2D DATA MODEL

**Signature**
Signature specifies the type of analyzer, the environment and any versioning information if available.

**Provenance**
Provenance specifies the type and value associated with each digital item's history. In the present context, creation date and modified date are associated with provenance.

**Context**
Context defines the metadata about digital item's existence. Usually, the mime-type, format, well formity, validity of the digital item are associated with context.

**Reference**
Reference holds the necessary information about digital items identity. This information entails the internal identifier and/or the identifier associated with the digital item.

**Fixity**
Fixity relates to any information regarding digital item that remains invariant to time. For example, checksum, size and printable stream are associated with fixity.

**Raw output**
This is defined to capture the analysis from the techniques as is, without any formatting.

**Type**
The type of elements defined for provenance, context, reference and fixity.

**Value**
The value associated with the corresponding type.

Figure 2 illustrates the data model used by the tool.

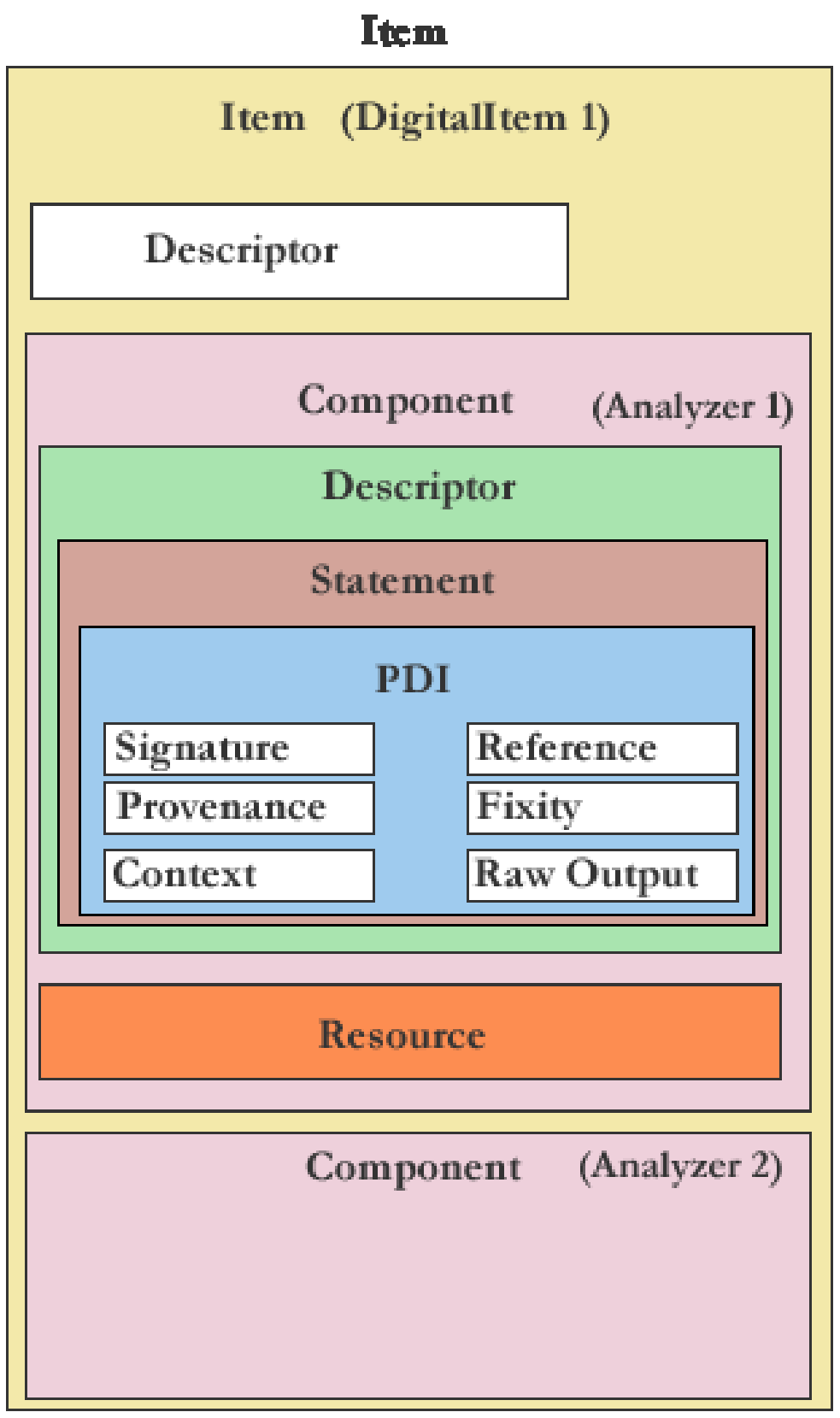

**Figure 2: D2D Data Model**

The D2D XML schema [21] and metadata captured inside D2D data model for the example digital item shown in table 2.The following table illustrates the usage of DID terms and D2D terms for the representation of analysis reports in DIDL.

**Table 14: Data Model & DIDL Mapping.**

| **Data Model** | **Enclosed elements** |
|---|---|
| Container | Descriptor, Item |
| Item | Component, Descriptor, Item |
| Component | Descriptor, Resource |
| Resource | Reference to Digital Item |
| Descriptor | Statement |
| Statement | PDI |
| PDI | Signature, Provenance, Context, Reference, Fixity, Raw output |
| Signature | Analyzer details |
| Provenance | Last Modified |
| Context | Format, mime type |
| Reference | Identifier, Internal Identifier |
| Fixity | CRC32, SHA-1, MD5, size |
| Raw output | Non-formatted output from analyzers |

## 5 D2D TOOL

The D2D tool is implemented in Java and provides a command line interface for performing various analysis techniques on a digital archive. The tool performs functions in three discrete phases namely ‘submission’, ‘analysis’, and ‘dissemination’. Figure 3 illustrates the activity flow inside the tool.

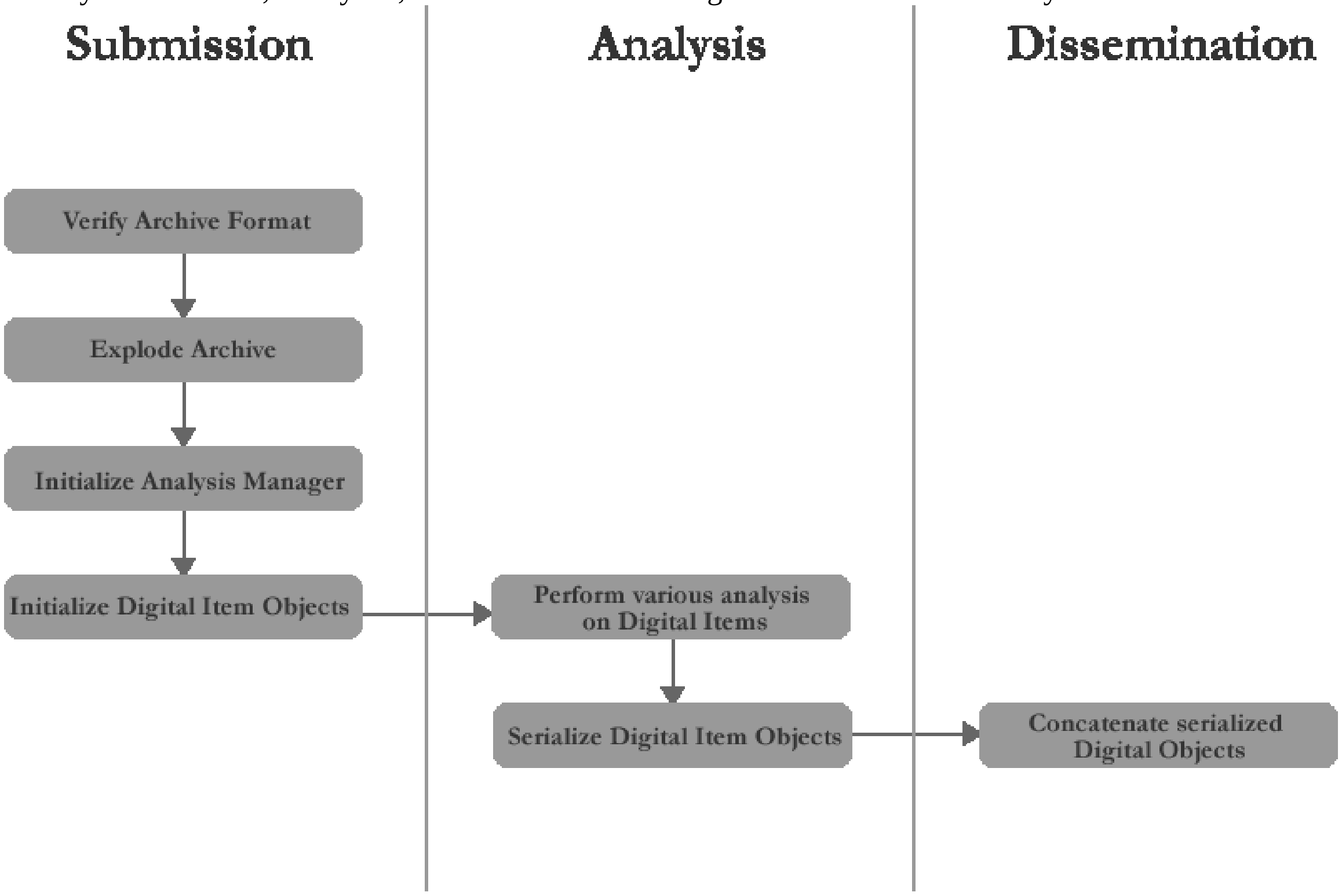

**Figure 3: Activity Diagram of D2D Tool**

**Submission**

`The ‘submission’ deals with the initialization of various entities for the tool to proceed with the analysis. The first step is to identify the submitted archive format. Since for each format, the tool has to explode the archive, it only supports a few archiving formats namely tar, zip, etc. Once, the archive is exploded, the tool initializes the ‘Analysis Manager’. This is a component inside the tool that applies various techniques over the digital items. The Analysis Manager is driven by the various ‘analysis techniques’ passed as a parameter to the tool. The final step in this phase is to initialize the digital item objects. Each digital item object is a run-time identity of the actual digital item grabbed from the archive. The initialization involves referencing the file system location, identifying the file size, assigning the identifier.

**Analysis**

The core functionality of the tool is implemented in this phase. The analysis manager performs the requested techniques over the digital items. The tool presently supports the various analysis techniques namely JHOVE, FRED, File, Checksum etc. The tool supports the addition of any new analysis techniques in a simple way. Once integrated with a new analysis technique, the manager automatically applies the new technique.

During this phase, the manager is also responsible for producing the timing report. The timing report is for individual analyzers as well as overall analysis. The section titled ‘Experiment Results' details about the timing reports for test cases.

The final step in this phase is to serialize the analyzed digital item. The serialization involves the creation of MPEG-21 DIDL components from the various analysis reports and synching them to disk. The DIDL components created conforms to the D2D data model and the XML schema.

**Dissemination**

The serialized digital items are read from disk and concatenated to produce the complete DIDL. Since each individual report is a valid XML file in itself, the DIDL produced is well-formed and valid. A sample DIDL is provided in Appendix A.

# 6 RUN-TIME PERFORMANCE

This section provides the timing report of the experiment analysis made on a sample digital archive. The sample archive for individual analysis timing report consisted of 148 files, and each file size ranging from 107 bytes to 3588344 bytes. The average file size was 41275.26 bytes, the largest file size is 3588344 bytes and the smallest file size is 107 bytes. It is important to note that the actual time is dependant on the machine it is analyzed; the results we present below were tested on a Intel(R) Xeon(TM), the details of which are given below:

CPU: Intel(R) Xeon(TM) CPU 3.20GHz (4)
OS: Linux 2.6.5-1 SMP
Total Memory: 1035660 kB

## 6.1 Individual Analysis Timing Report

**File Analysis**

Figure 4 shows the time required to run “file” vs. the file sizes of digital items. It is interesting to note that the time it takes to analyze is consistent regardless of the size of the digital item. It is attributed to the fact that, ‘file’ analyzes based on the first few bytes of the content and hence, the size of the digital item is often not an important factor for the analysis. The consistent fluctuations seen in the active zone from 20ms to 120ms is explained as follows. The language test that the file performs is usually terminated when the test is successful. In the tests conducted, the maximum time that it takes to perform the ‘file’ analysis including the language test is just above 120ms. The maximum size of a digital item on which the analysis is done (not shown here) is 29.5MB and it took 18ms.

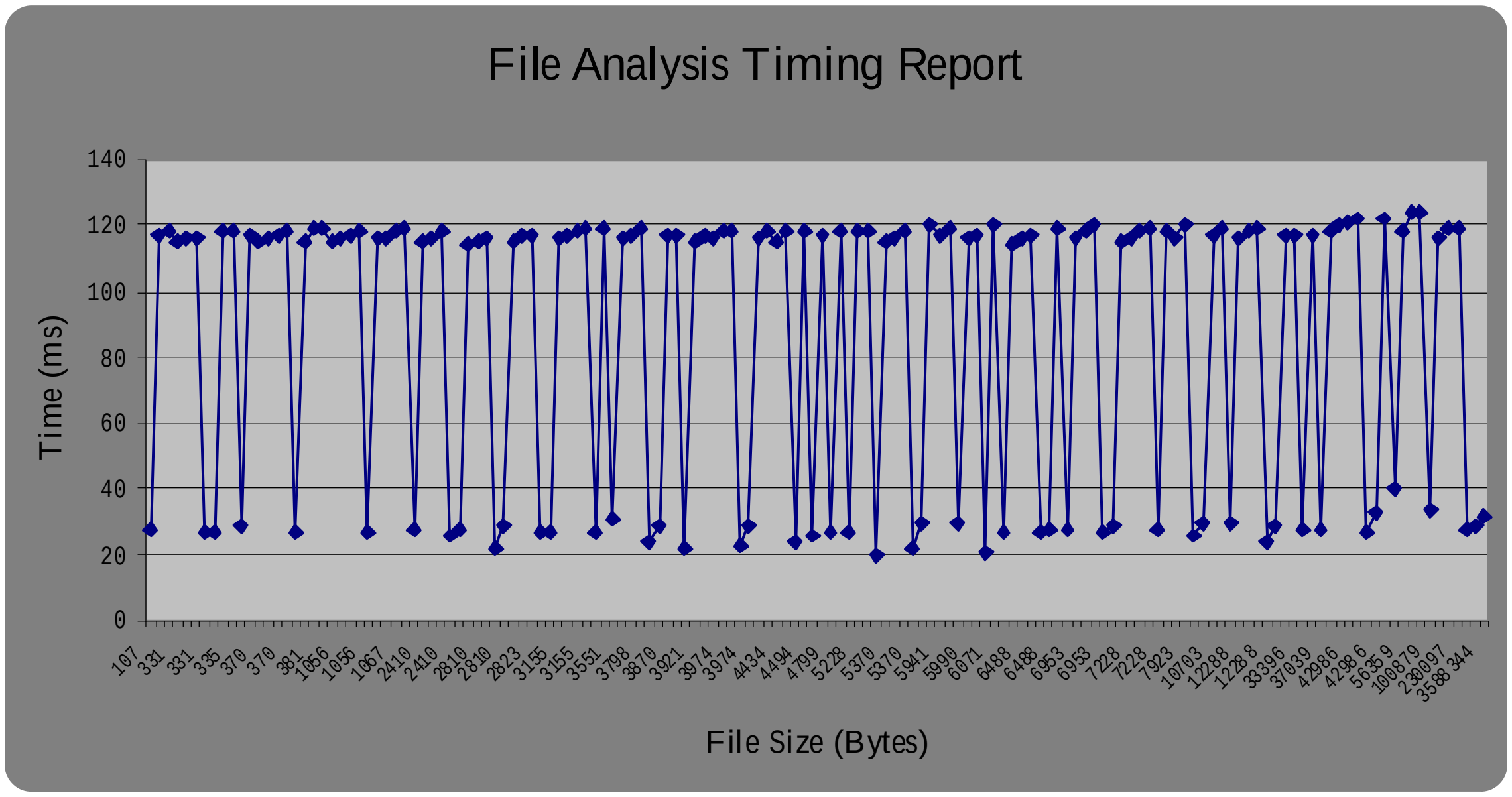

**Figure 4. Time Required to Run "file".**

**Raw characters analysis**

As mentioned above, the 'strings' command is used to find the raw characters from the digital items. Figure 5 shows the time required to run "strings" vs. file size on the x-axis and time in milliseconds on the y-axis. It is interesting to notice that the time it takes to perform the 'strings' analysis is consistent until a threshold, after which it peaks out. The largest file takes the largest time to be processed by 'strings' command. The average time for the string command to process on the files is consistent (approximately 50ms), with the exception of the last file because of its increase in size. The peaking out of time is expected, but it is interesting to note that the time it took until the threshold is pretty consistent. One of the factors could be the initiation of the program to perform the task might be a huge factor until the threshold is reached.

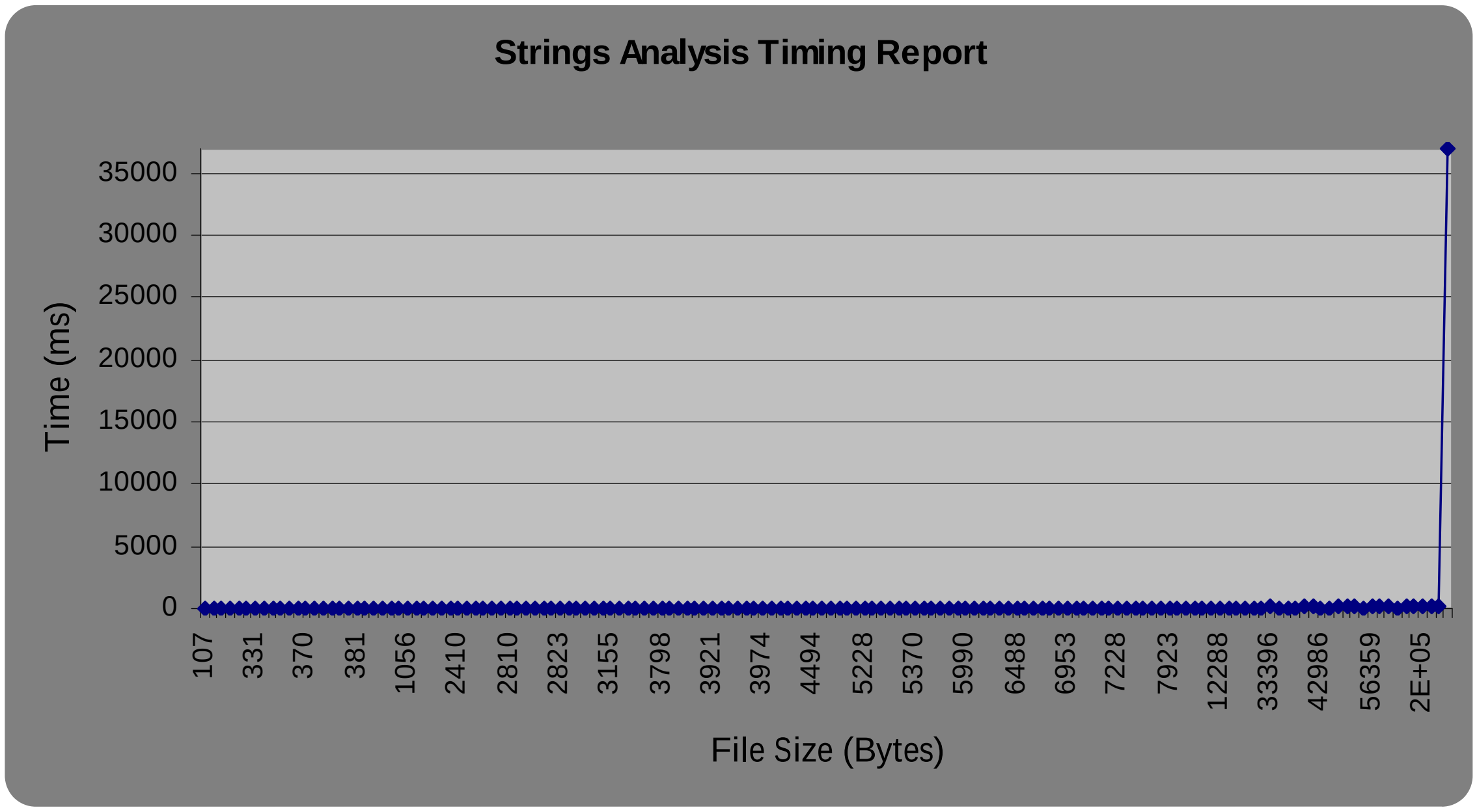

**Figure 5. Time Required to Run "strings".**

**Checksum Analysis**

Figure 6 shows the time required to compute checksums as carried out using the Java standard development kit. The checksums performed involves the computation of both MD5 and SHA-1. It is interesting to see that the time it takes to perform is consistent until a certain threshold and later it peaked. The reason for doing so may be attributed to the fact that the initialization time is overriding the computational time until the threshold is reached. For file sizes over the threshold, the initialization time is less than the actual computational time.

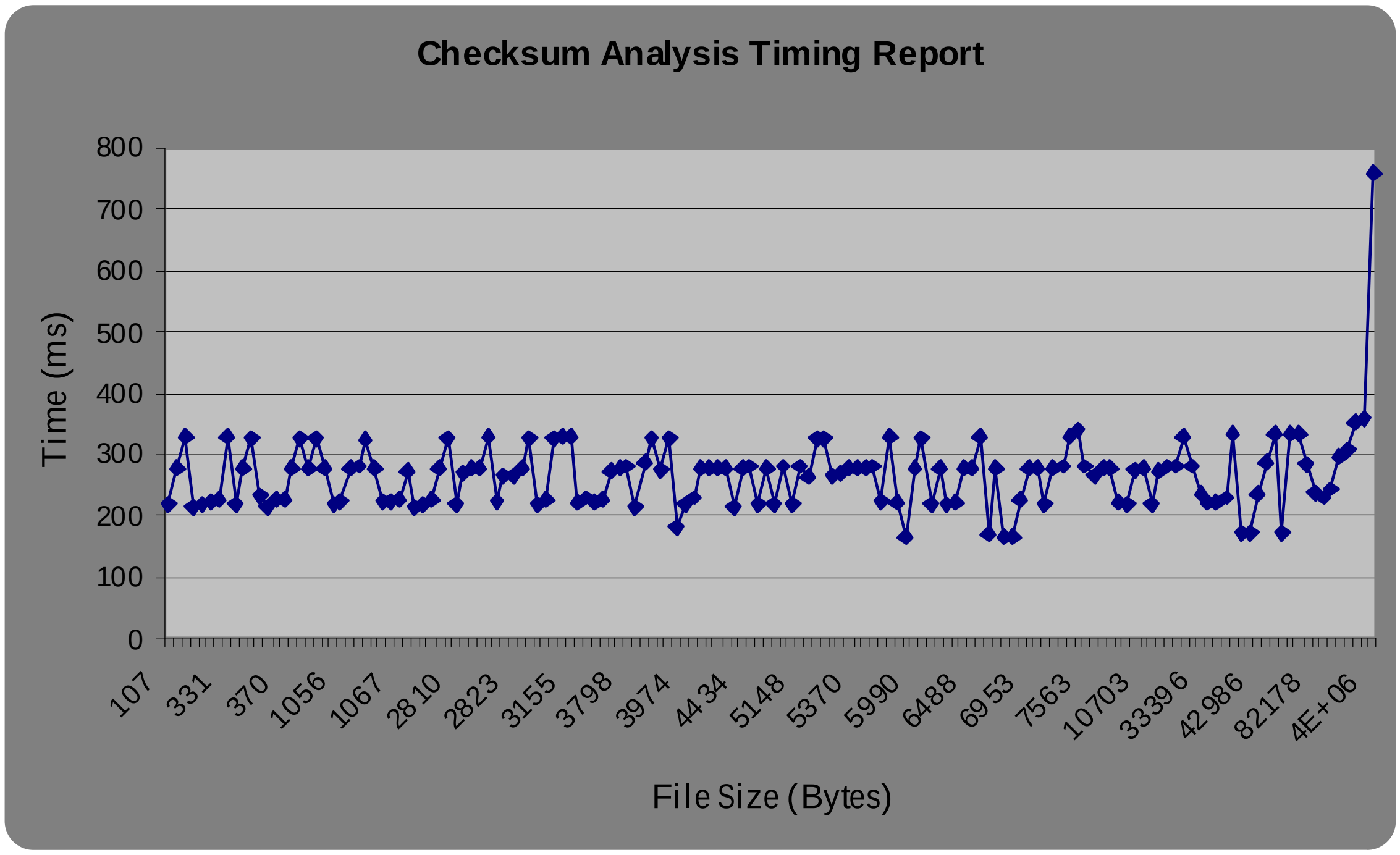

**Figure 6. Time Required to Compute Message-Digests.**

**JHOVE analysis**

JHOVE has the most consistent performance; the average time taken is approximately between 500ms and 1000ms. The rare peaks in time are when JHOVE encountered the XML files. JHOVE performs the validation of XML, and it is expected to be time consuming. Figure 7 shows the time required to run JHOVE for various file sizes.

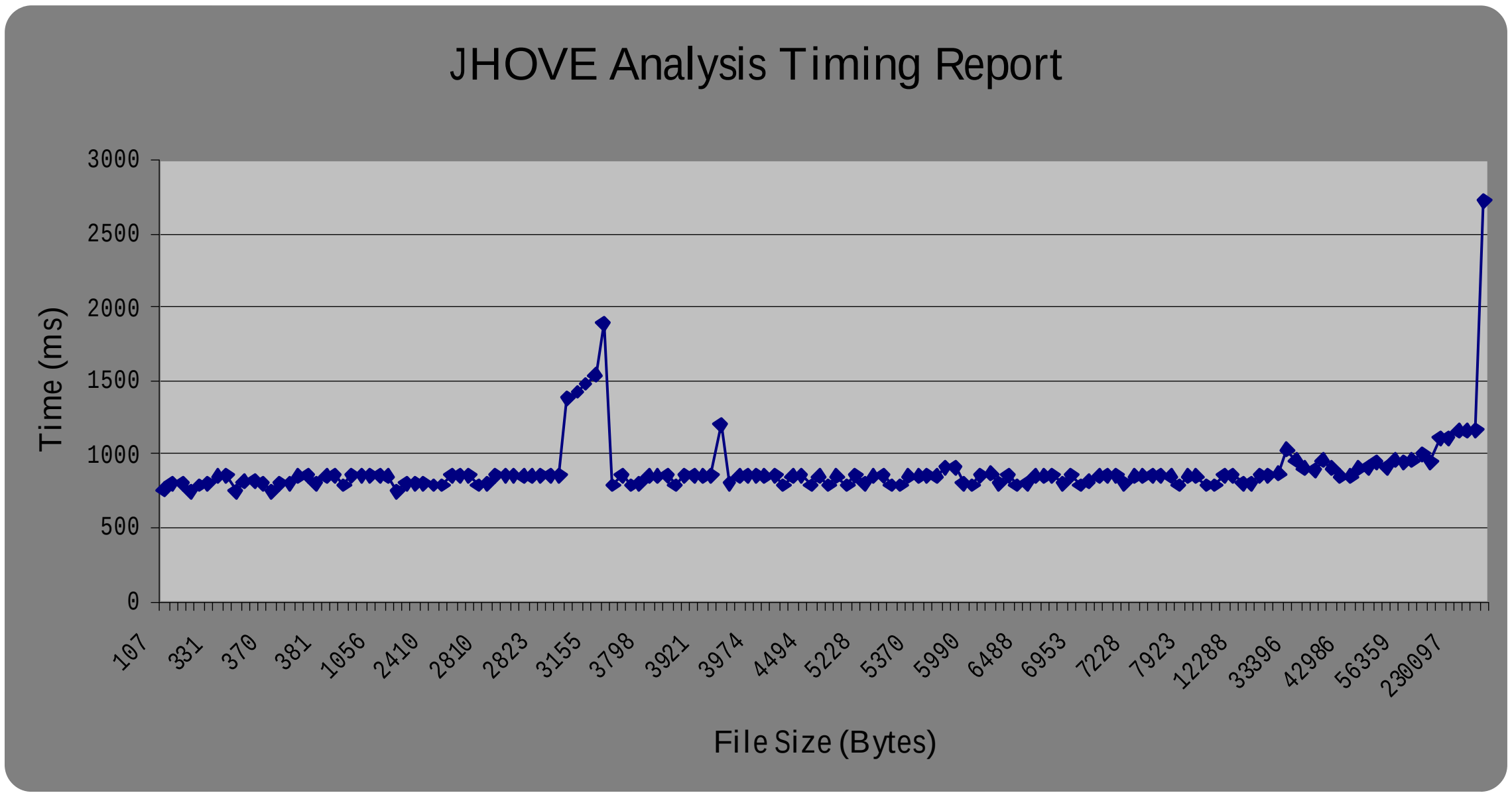

**Figure 7. Time Required to Run JHOVE.**

## Aggregate Timing

Figure 8 plots Archive size versus Time. The labels on each plotted point represent the number of files in the archive. The graphs clearly suggest that the number of files is the primary factor determining the time rather than the archive size.

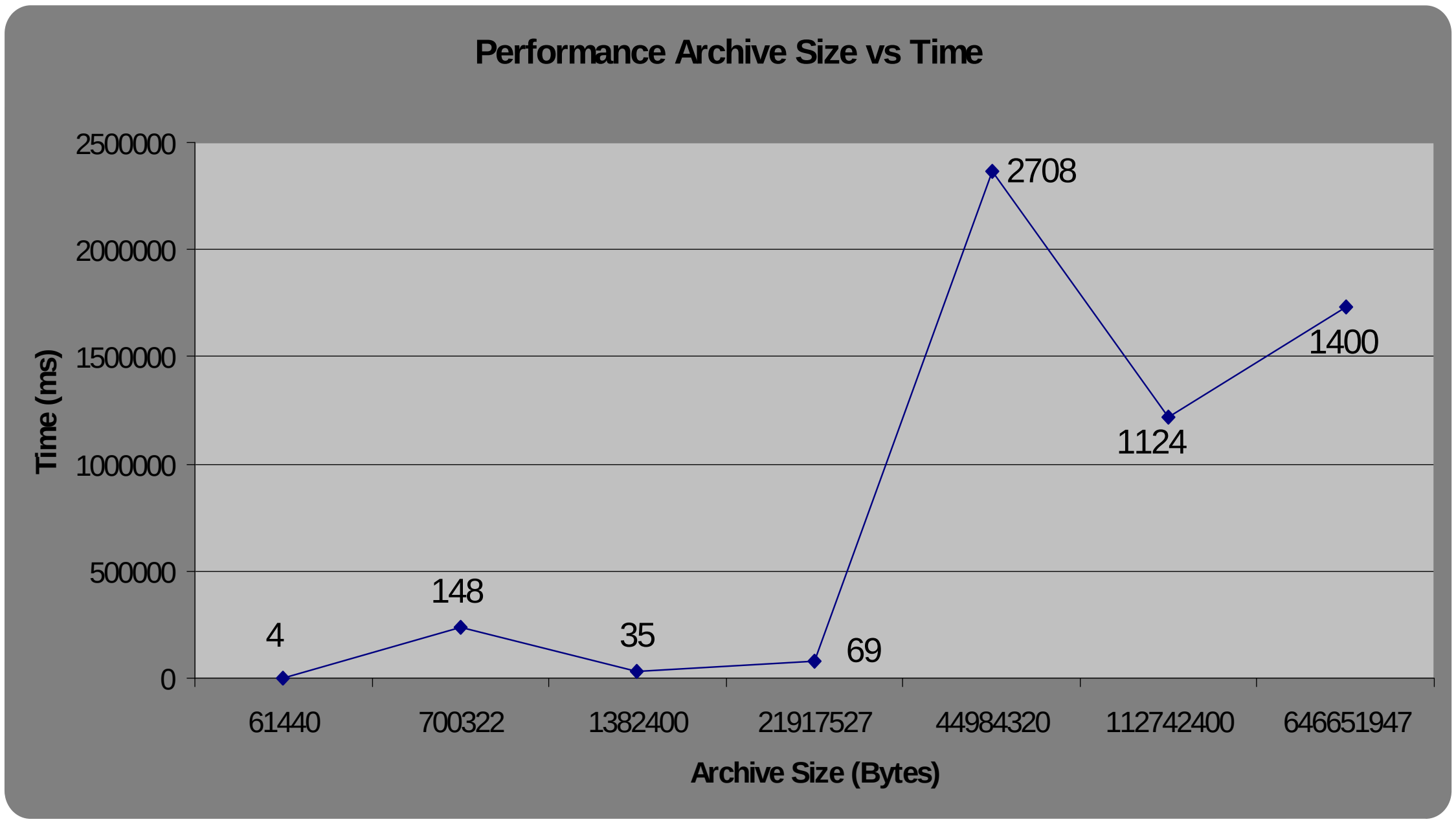

**Figure 8. Aggregate Timings of Archive Processing.**

# 7 CONCLUSIONS

Because it is difficult to realistically simulate the passage of time, research in digital preservation is difficult to evaluate. Our approach can be characterized as providing a framework to aggregate the output of a variety of digital content introspection utilities. We have used a variety of commonly available tools, but the arrangement of the tools is extensible and can be adapted to a particular preference or profile. D2D is a proof-of-concept tool for the community participating in this effort. The timing analysis performed helps the community to understand the amount of computational resources involved in this study. Thus, this project not only provides a tool that can be used for analysis, but also provided some timing reports to demonstrate feasibility and provide benchmarks for future tools.

## APPENDIX A – Sample D2D Output

```
  <?xml version="1.0" encoding="UTF-8" ?>
- <didl:DIDL xmlns:didl="urn:mpeg:mpeg21:2002:02-DIDL-NS"
xmlns:daap="http://www.cs.odu.edu/DAAP"
xmlns:xsi="http://www.w3.org/2001/XMLSchema-instance"
xsi:schemaLocation="http://www.cs.odu.edu/DAAP
http://www.cs.odu.edu/%7Egmanepal/chucchi/DAAP.xsd urn:mpeg:mpeg21:2002:02-
DIDL-NS http://www.cs.odu.edu/%7Egmanepal/chucchi/didl.xsd">
- <didl:Container>
- <didl:Descriptor>
  <didl:Statement mimeType="text/plain; charset=UTF-8" />
  </didl:Descriptor>
- <didl:Item>
- <didl:Component>
  <didl:Resource mimeType="text/plain; charset=US-ASCII" ref="test.tar.gz" />
  </didl:Component>
- <didl:Item>
- <didl:Component>
- <didl:Descriptor>
- <didl:Statement mimeType="text/xml; charset=UTF-8">
- <daap:PDI>
  <daap:signature>file-4.07 magic file from /usr/share/file/magic</daap:signature>
- <daap:context>
  <daap:type>mimeType</daap:type>
  <daap:value>text/plain; charset=us-ascii</daap:value>
  </daap:context>
- <daap:reference>
  <daap:type>identifier</daap:type>
  <daap:value>100.700/5EFFE0D18D8CB5910EECEE5D31337A40</daap:value>
  </daap:reference>
- <daap:reference>
  <daap:type>internalIdengztifier</daap:type>
  <daap:value>5EFFE0D18D8CB5910EECEE5D31337A40.0</daap:value>
  </daap:reference>
  <daap:rawOutput>ASCII text, with very long lines</daap:rawOutput>
  </daap:PDI>
  </didl:Statement>
  </didl:Descriptor>
  <didl:Resource mimeType="text/plain; charset:utf8">file-4.07 magic file from
/usr/share/file/magic</didl:Resource>
  </didl:Component>
- <didl:Component>
- <didl:Descriptor>
- <didl:Statement mimeType="text/xml; charset=UTF-8">
- <daap:PDI>
  <daap:signature>java version "1.4.2_05" Java(TM) 2 Runtime Environment, Standard
Edition (build 1.4.2_05-b04) Java HotSpot(TM) Client VM (build 1.4.2_05-b04, mixed
mode)</daap:signature>
- <daap:reference>
  <daap:type>identifier</daap:type>
  <daap:value>100.700/5EFFE0D18D8CB5910EECEE5D31337A40</daap:value>
  </daap:reference>
- <daap:reference>
  <daap:type>internalIdentifier</daap:type>
  <daap:value>5EFFE0D18D8CB5910EECEE5D31337A40.0</daap:value>
  </daap:reference>
- <daap:fixity>
  <daap:type>SHA</daap:type>
  <daap:value>1E18C123927027C0EE5F8DA121B5CB5ADA62710A</daap:value>
  </daap:fixity>
- <daap:fixity>
```

<daap:type>MD5</daap:type>
<daap:value>9CCA20DA7203E90EE8FBEF6C53C3AB1D</daap:value>
</daap:fixity>
<daap:rawOutput>SHA:1E18C123927027C0EE5F8DA121B5CB5ADA62710A MD5:9CCA20DA7203E90EE8FBEF6C53C3AB1D</daap:rawOutput>
</daap:PDI>
</didl:Statement>
</didl:Descriptor>
<didl:Resource mimeType="text/plain; charset:utf8">java version "1.4.2_05" Java(TM) 2 Runtime Environment, Standard Edition (build 1.4.2_05-b04) Java HotSpot(TM) Client VM (build 1.4.2_05-b04, mixed mode)</didl:Resource>
</didl:Component>
- <didl:Component>
- <didl:Descriptor>
- <didl:Statement mimeType="text/xml; charset=UTF-8">
- <daap:PDI>
<daap:signature>Jhove (Rel. 1.0 (beta 2), 2004-07-19) Date: 2005-12-02 11:52:18 EST App: Module: BYTESTREAM 1.0 OutputHandler: TEXT 1.1 OutputHandler: XML 1.1 Usage: java Jhove [-c conf] [-m module [-p param]] [-h handler] [-e encoding] [-H handler] [-o output] [-x saxclass] [-t tempdir] [-b bufsize] [[-krs] file-or-uri] Rights: Copyright 2003-2004 by JSTOR and the President and Fellows of Harvard College. Released under the GNU Lesser General Public License.</daap:signature>
- <daap:provenance>
<daap:type>lastModified</daap:type>
<daap:value>2005-11-27T12:15:43-05:00</daap:value>
</daap:provenance>
- <daap:context>
<daap:type>mimeType</daap:type>
<daap:value>text/plain; charset=US-ASCII</daap:value>
</daap:context>
- <daap:context>
<daap:type>status</daap:type>
<daap:value>Well-formed and valid</daap:value>
</daap:context>
- <daap:context>
<daap:type>format</daap:type>
<daap:value>ASCII</daap:value>
</daap:context>
- <daap:reference>
<daap:type>identifier</daap:type>
<daap:value>100.700/5EFFE0D18D8CB5910EECEE5D31337A40</daap:value>
</daap:reference>
- <daap:reference>
<daap:type>internalIdentifier</daap:type>
<daap:value>5EFFE0D18D8CB5910EECEE5D31337A40.0</daap:value>
</daap:reference>
- <daap:fixity>
<daap:type>size</daap:type>
<daap:value>6968</daap:value>
</daap:fixity>
- <daap:fixity>
<daap:type>SHA-1</daap:type>
<daap:value>1e18c123927027c0ee5f8da121b5cb5ada6271a</daap:value>
</daap:fixity>
- <daap:fixity>
<daap:type>MD5</daap:type>
<daap:value>9cca20da723e9ee8fbef6c53c3ab1d</daap:value>
</daap:fixity>
- <daap:fixity>
<daap:type>CRC32</daap:type>
<daap:value>a0a74972</daap:value>
</daap:fixity>

```
  <daap:rawOutput><?xml version="1.0" encoding="UTF-8"?> <jhove
xmlns:xsi="http://www.w3.org/2001/XMLSchema-instance"
xmlns="http://hul.harvard.edu/ois/xml/ns/jhove"
xsi:schemaLocation="http://hul.harvard.edu/ois/xml/ns/jhove
http://hul.harvard.edu/ois/xml/xsd/jhove/jhove.xsd" name="Jhove" release="1.0 (beta
2)" date="2004-07-19"> <date>2005-12-02T11:52:17-05:00</date> <repInfo
uri="test/error.xml"> <reportingModule release="1.0" date="2004-05-05">ASCII-
hul</reportingModule> <lastModified>2005-11-27T12:15:43-05:00</lastModified>
<size>6968</size> <format>ASCII</format> <status>Well-formed and valid</status>
<mimeType>text/plain; charset=US-ASCII</mimeType> <properties> <property>
<name>ASCIIMetadata</name> <values arity="List" type="Property"> <property>
<name>LineEndings</name> <values arity="List" type="String"> <value>LF</value>
</values> </property> </values> </property> </properties> <checksums> <checksum
type="CRC32">a0a74972</checksum> <checksum
type="MD5">9cca20da723e9ee8fbef6c53c3ab1d</checksum> <checksum
type="SHA-1">1e18c123927027c0ee5f8da121b5cb5ada6271a</checksum>
</checksums> </repInfo> </jhove></daap:rawOutput>
  </daap:PDI>
  </didl:Statement>
  </didl:Descriptor>
  <didl:Resource mimeType="text/plain; charset:utf8">Jhove (Rel. 1.0 (beta 2), 2004-07-19)
Date: 2005-12-02 11:52:18 EST App: Module: BYTESTREAM 1.0 OutputHandler: TEXT 1.1
OutputHandler: XML 1.1 Usage: java Jhove [-c conf] [-m module [-p param]] [-h handler]
[-e encoding] [-H handler] [-o output] [-x saxclass] [-t tempdir] [-b bufsize] [[-krs] file-
or-uri] Rights: Copyright 2003-2004 by JSTOR and the President and Fellows of Harvard
College. Released under the GNU Lesser General Public License.</didl:Resource>
  </didl:Component>
- <didl:Component>
- <didl:Descriptor>
- <didl:Statement mimeType="text/xml; charset=UTF-8">
- <daap:PDI>
  <daap:signature>GNU strings 2.15.90.0.3 20040415 Copyright 2004 Free Software
Foundation, Inc. This program is free software; you may redistribute it under the terms of
the GNU General Public License. This program has absolutely no
warranty.</daap:signature>
- <daap:reference>
  <daap:type>identifier</daap:type>
  <daap:value>100.700/5EFFE0D18D8CB5910EECEE5D31337A40</daap:value>
  </daap:reference>
- <daap:reference>
  <daap:type>internalIdentifier</daap:type>
  <daap:value>5EFFE0D18D8CB5910EECEE5D31337A40.0</daap:value>
  </daap:reference>
- <daap:fixity>
  <daap:type>rawCharacters</daap:type>
  <daap:value><didl:Component> <didl:Descriptor> <didl:Statement
mimeType="text/xml; charset=UTF-8"> <dc:identifier
xmlns:dc="http://purl.org/dc/elements/1.1/"
xmlns:xsi="http://www.w3.org/2001/XMLSchema-instance"
xsi:schemaLocation="http://purl.org/dc/elements/1.1/
http://dublincore.org/schemas/xmls/simpledc20021212.xsd"></dc:identifier>
<dc:source xmlns:dc="http://purl.org/dc/elements/1.1/"
xmlns:xsi="http://www.w3.org/2001/XMLSchema-instance"
xsi:schemaLocation="http://purl.org/dc/elements/1.1/
http://dublincore.org/schemas/xmls/simpledc20021212.xsd">Track_11.cdda</dc:sourc
e> </didl:Statement> </didl:Descriptor> <didl:Descriptor> <didl:Statement
mimeType="text/xml; charset=UTF-8"> <dc:creator
xmlns:dc="http://purl.org/dc/elements/1.1/"
xmlns:xsi="http://www.w3.org/2001/XMLSchema-instance"
xsi:schemaLocation="http://purl.org/dc/elements/1.1/
http://dublincore.org/schemas/xmls/simpledc20021212.xsd">MIMEtype</dc:creator>
<dc:description xmlns:dc="http://purl.org/dc/elements/1.1/"
xmlns:xsi="http://www.w3.org/2001/XMLSchema-instance"
xsi:schemaLocation="http://purl.org/dc/elements/1.1/
http://dublincore.org/schemas/xmls/simpledc20021212.xsd">
```

application/aiff</dc:description> </didl:Statement> </didl:Descriptor>
<didl:Descriptor> <didl:Statement mimeType="text/xml; charset=UTF-8"> <dc:creator
xmlns:dc="http://purl.org/dc/elements/1.1/"
xmlns:xsi="http://www.w3.org/2001/XMLSchema-instance"
xsi:schemaLocation="http://purl.org/dc/elements/1.1/
http://dublincore.org/schemas/xmls/simpledc20021212.xsd">Jhove (Rel. 1.0 (beta 2),
2004-07-19) </dc:creator> <dc:description
xmlns:dc="http://purl.org/dc/elements/1.1/"
xmlns:xsi="http://www.w3.org/2001/XMLSchema-instance"
xsi:schemaLocation="http://purl.org/dc/elements/1.1/
http://dublincore.org/schemas/xmls/simpledc20021212.xsd"> Jhove (Rel. 1.0 (beta 2),
2004-07-19) Date: 2004-12-29 23:02:15 EST RepresentationInformation:
websites/chnm/september11/REPOSITORY/CONTRIBUTORS/sonic_mem_preview/Track_
11.cdda ReportingModule: AIFF-hul, Rel. 1.0 (2004-07-12) LastModified: 2004-01-27
13:05:15 EST Size: 101361872 Format: AIFF Status: Well-formed and valid MIMEtype:
application/aiff Profile: AIFF-C AIFFMetadata: AESAudioMetadata: AnalogDigitalFlag:
FILE_DIGITAL SchemaVersion: 1.02b Format: AIFF-C SpecificationVersion: Draft
1991-08-26 AudioDataEncoding: PCM PrimaryIdentifier:
websites/chnm/september11/REPOSITORY/CONTRIBUTORS/sonic_mem_preview/Track_
11.cdda IdentifierType: FILE_NAME FormatList: FormatRegion: BitDepth: 16 SampleRate:
44100.0 BitDepth: 16 NumChannels: 2 FormatVersion: Wed May 23 15:40:00 EDT 1990
SampleFrames: 25340448 CompressionType: sowt CompressionName: SoundData: Offset:
8 BlockSize: 0 DataLength: 101361800</dc:description> </didl:Statement>
</didl:Descriptor> <didl:Descriptor> <didl:Statement mimeType="text/xml;
charset=UTF-8"> <dc:creator xmlns:dc="http://purl.org/dc/elements/1.1/"
xmlns:xsi="http://www.w3.org/2001/XMLSchema-instance"
xsi:schemaLocation="http://purl.org/dc/elements/1.1/
http://dublincore.org/schemas/xmls/simpledc20021212.xsd">file-4.07 magic file from
/usr/share/file/magic </dc:creator> <dc:description
xmlns:dc="http://purl.org/dc/elements/1.1/"
xmlns:xsi="http://www.w3.org/2001/XMLSchema-instance"
xsi:schemaLocation="http://purl.org/dc/elements/1.1/
http://dublincore.org/schemas/xmls/simpledc20021212.xsd"> IFF data
</dc:description> </didl:Statement> </didl:Descriptor> <didl:Descriptor>
<didl:Statement mimeType="text/xml; charset=UTF-8"> <dc:creator
xmlns:dc="http://purl.org/dc/elements/1.1/"
xmlns:xsi="http://www.w3.org/2001/XMLSchema-instance"
xsi:schemaLocation="http://purl.org/dc/elements/1.1/
http://dublincore.org/schemas/xmls/simpledc20021212.xsd">Fred Info
URI</dc:creator> <dc:description xmlns:dc="http://purl.org/dc/elements/1.1/"
xmlns:xsi="http://www.w3.org/2001XMLSchema-instance"
xsi:schemaLocation="http://purl.org/dc/elements/1.1/
http://dublincore.org/schemas/xmls/simpledc2002121.xsd">
info:gdfr/fred/f/aiff</dc:description> </didl:Statement> </didl:Descriptor>
<didl:Resource m<didl:Component> <didl:Descriptor> <didl:Statement
mimeType="text/xml; charset=UTF-8"> <dc:identifier
xmlns:dc="http://purl.org/dc/elements/1.1/"
xmlns:xsi="http://www.w3.org/2001/XMLSchema-instance"
xsi:schemaLocation="http://purl.org/dc/elements/1.1/
http://dublincore.org/schemas/xmls/simpledc20021212.xsd">
6a18d2726cd6340b076a02cd8738d63</dc:identifier> <dc:source
xmlns:dc="http://purl.org/dc/elements/1.1/"
xmlns:xsi="http://www.w3.org/2001/XMLSchema-instance"
xsi:schemaLocation="http://purl.org/dc/elements/1.1/
http://dublincore.org/schemas/xmls/simpledc20021212.xsd">WTC009.jpg</dc:source>
</didl:Statement> </didl:Descriptor> <didl:Descriptor> <didl:Statement
mimeType="text/xml; charset=UTF-8"> <dc:creator
xmlns:dc="http://purl.org/dc/elements/1.1/"
xmlns:xsi="http://www.w3.org/2001/XMLSchema-instance"
xsi:schemaLocation="http://purl.org/dc/elements/1.1/
http://dublincore.org/schemas/xmls/simpledc20021212.xsd">MIMEtype</dc:creator>
<dc:description xmlns:dc="http://purl.org/dc/elements/1.1/"
xmlns:xsi="http://www.w3.org/2001/XMLSchema-instance"
xsi:schemaLocation="http://purl.org/dc/elements/1.1/
http://dublincore.org/schemas/xmls/simpledc20021212.xsd">
image/jpeg</dc:description> </didl:Statement> </didl:Descriptor> <didl:Descriptor>

<didl:Statement mimeType="text/xml; charset=UTF-8"> <dc:creator
xmlns:dc="http://purl.org/dc/elements/1.1/"
xmlns:xsi="http://www.w3.org/2001/XMLSchema-instance"
xsi:schemaLocation="http://purl.org/dc/elements/1.1/
http://dublincore.org/schemas/xmls/simpledc20021212.xsd">Jhove (Rel. 1.0 (beta 2),
2004-07-19) </dc:creator> <dc:description
xmlns:dc="http://purl.org/dc/elements/1.1/"
xmlns:xsi="http://www.w3.org/2001/XMLSchema-instance"
xsi:schemaLocation="http://purl.org/dc/elements/1.1/
http://dublincore.org/schemas/xmls/simpledc20021212.xsd"> Jhove (Rel. 1.0 (beta 2),
2004-07-19) Date: 2004-12-26 22:08:25 EST RepresentationInformation:
websites/chnm/september11/REPOSITORY/CONTRIBUTORS/steve_levine/WTC009.jpg
ReportingModule: JPEG-hul, Rel. 1.0 (2004-06-23) LastModified: 2004-01-27 13:05:25 EST
Size: 729814 Format: JPEG Version: 1.02 Status: Well-formed and valid MIMEtype:
image/jpeg Profile: JFIF JPEGMetadata: CompressionType: Huffman coding, Baseline DCT
Images: Image: NisoImageMetadata: MIMEType: image/jpeg ByteOrder: big-endian
CompressionScheme: JPEG ColorSpace: YCbCr SamplingFrequencyUnit: inch ImageWidth:
1024 ImageLength: 841 BitsPerSample: 8 SamplesPerPixel: 3 RestartInterval: 128 Scans:
1 QuantizationTables: QuantizationTable: Precision: 8-bit DestinationIdentifier: 0
ApplicationSegments: APP0, APP13, APP14 Checksum: c176826e Type: CRC32 Checksum:
6a18d2726cd6340b076a02cd8738d63 Type: MD5 Checksum:
d3d99e9f5162169c295370cede6fbbc0c4f5b2d2 Type:
SHA-1</dc:description></daap:value>
</daap:fixity>
<daap:rawOutput><didl:Component> <didl:Descriptor> <didl:Statement
mimeType="text/xml; charset=UTF-8"> <dc:identifier
xmlns:dc="http://purl.org/dc/elements/1.1/"
xmlns:xsi="http://www.w3.org/2001/XMLSchema-instance"
xsi:schemaLocation="http://purl.org/dc/elements/1.1/
http://dublincore.org/schemas/xmls/simpledc20021212.xsd"></dc:identifier>
<dc:source xmlns:dc="http://purl.org/dc/elements/1.1/"
xmlns:xsi="http://www.w3.org/2001/XMLSchema-instance"
xsi:schemaLocation="http://purl.org/dc/elements/1.1/
http://dublincore.org/schemas/xmls/simpledc20021212.xsd">Track_11.cdda</dc:sourc
e> </didl:Statement> </didl:Descriptor> <didl:Descriptor> <didl:Statement
mimeType="text/xml; charset=UTF-8"> <dc:creator
xmlns:dc="http://purl.org/dc/elements/1.1/"
xmlns:xsi="http://www.w3.org/2001/XMLSchema-instance"
xsi:schemaLocation="http://purl.org/dc/elements/1.1/
http://dublincore.org/schemas/xmls/simpledc20021212.xsd">MIMEtype</dc:creator>
<dc:description xmlns:dc="http://purl.org/dc/elements/1.1/"
xmlns:xsi="http://www.w3.org/2001/XMLSchema-instance"
xsi:schemaLocation="http://purl.org/dc/elements/1.1/
http://dublincore.org/schemas/xmls/simpledc20021212.xsd">
application/aiff</dc:description> </didl:Statement> </didl:Descriptor>
<didl:Descriptor> <didl:Statement mimeType="text/xml; charset=UTF-8"> <dc:creator
xmlns:dc="http://purl.org/dc/elements/1.1/"
xmlns:xsi="http://www.w3.org/2001/XMLSchema-instance"
xsi:schemaLocation="http://purl.org/dc/elements/1.1/
http://dublincore.org/schemas/xmls/simpledc20021212.xsd">Jhove (Rel. 1.0 (beta 2),
2004-07-19) </dc:creator> <dc:description
xmlns:dc="http://purl.org/dc/elements/1.1/"
xmlns:xsi="http://www.w3.org/2001/XMLSchema-instance"
xsi:schemaLocation="http://purl.org/dc/elements/1.1/
http://dublincore.org/schemas/xmls/simpledc20021212.xsd"> Jhove (Rel. 1.0 (beta 2),
2004-07-19) Date: 2004-12-29 23:02:15 EST RepresentationInformation:
websites/chnm/september11/REPOSITORY/CONTRIBUTORS/sonic_mem_preview/Track_
11.cdda ReportingModule: AIFF-hul, Rel. 1.0 (2004-07-12) LastModified: 2004-01-27
13:05:15 EST Size: 101361872 Format: AIFF Status: Well-formed and valid MIMEtype:
application/aiff Profile: AIFF-C AIFFMetadata: AESAudioMetadata: AnalogDigitalFlag:
FILE_DIGITAL SchemaVersion: 1.02b Format: AIFF-C SpecificationVersion: Draft
1991-08-26 AudioDataEncoding: PCM PrimaryIdentifier:
websites/chnm/september11/REPOSITORY/CONTRIBUTORS/sonic_mem_preview/Track_
11.cdda IdentifierType: FILE_NAME FormatList: FormatRegion: BitDepth: 16 SampleRate:
44100.0 BitDepth: 16 NumChannels: 2 FormatVersion: Wed May 23 15:40:00 EDT 1990
SampleFrames: 25340448 CompressionType: sowt CompressionName: SoundData: Offset:

8 BlockSize: 0 DataLength: 101361800</dc:description> </didl:Statement>
</didl:Descriptor> <didl:Descriptor> <didl:Statement mimeType="text/xml;
charset=UTF-8"> <dc:creator xmlns:dc="http://purl.org/dc/elements/1.1/"
xmlns:xsi="http://www.w3.org/2001/XMLSchema-instance"
xsi:schemaLocation="http://purl.org/dc/elements/1.1/
http://dublincore.org/schemas/xmls/simpledc20021212.xsd">file-4.07 magic file from
/usr/share/file/magic </dc:creator> <dc:description
xmlns:dc="http://purl.org/dc/elements/1.1/"
xmlns:xsi="http://www.w3.org/2001/XMLSchema-instance"
xsi:schemaLocation="http://purl.org/dc/elements/1.1/
http://dublincore.org/schemas/xmls/simpledc20021212.xsd"> IFF data
</dc:description> </didl:Statement> </didl:Descriptor> <didl:Descriptor>
<didl:Statement mimeType="text/xml; charset=UTF-8"> <dc:creator
xmlns:dc="http://purl.org/dc/elements/1.1/"
xmlns:xsi="http://www.w3.org/2001/XMLSchema-instance"
xsi:schemaLocation="http://purl.org/dc/elements/1.1/
http://dublincore.org/schemas/xmls/simpledc20021212.xsd">Fred Info
URI</dc:creator> <dc:description xmlns:dc="http://purl.org/dc/elements/1.1/"
xmlns:xsi="http://www.w3.org/2001XMLSchema-instance"
xsi:schemaLocation="http://purl.org/dc/elements/1.1/
http://dublincore.org/schemas/xmls/simpledc2002121.xsd">
info:gdfr/fred/f/aiff</dc:description> </didl:Statement> </didl:Descriptor>
<didl:Resource m<didl:Component> <didl:Descriptor> <didl:Statement
mimeType="text/xml; charset=UTF-8"> <dc:identifier
xmlns:dc="http://purl.org/dc/elements/1.1/"
xmlns:xsi="http://www.w3.org/2001/XMLSchema-instance"
xsi:schemaLocation="http://purl.org/dc/elements/1.1/
http://dublincore.org/schemas/xmls/simpledc20021212.xsd">
6a18d2726cd6340b076a02cd8738d63</dc:identifier> <dc:source
xmlns:dc="http://purl.org/dc/elements/1.1/"
xmlns:xsi="http://www.w3.org/2001/XMLSchema-instance"
xsi:schemaLocation="http://purl.org/dc/elements/1.1/
http://dublincore.org/schemas/xmls/simpledc20021212.xsd">WTC009.jpg</dc:source>
</didl:Statement> </didl:Descriptor> <didl:Descriptor> <didl:Statement
mimeType="text/xml; charset=UTF-8"> <dc:creator
xmlns:dc="http://purl.org/dc/elements/1.1/"
xmlns:xsi="http://www.w3.org/2001/XMLSchema-instance"
xsi:schemaLocation="http://purl.org/dc/elements/1.1/
http://dublincore.org/schemas/xmls/simpledc20021212.xsd">MIMEtype</dc:creator>
<dc:description xmlns:dc="http://purl.org/dc/elements/1.1/"
xmlns:xsi="http://www.w3.org/2001/XMLSchema-instance"
xsi:schemaLocation="http://purl.org/dc/elements/1.1/
http://dublincore.org/schemas/xmls/simpledc20021212.xsd">
image/jpeg</dc:description> </didl:Statement> </didl:Descriptor> <didl:Descriptor>
<didl:Statement mimeType="text/xml; charset=UTF-8"> <dc:creator
xmlns:dc="http://purl.org/dc/elements/1.1/"
xmlns:xsi="http://www.w3.org/2001/XMLSchema-instance"
xsi:schemaLocation="http://purl.org/dc/elements/1.1/
http://dublincore.org/schemas/xmls/simpledc20021212.xsd">Jhove (Rel. 1.0 (beta 2),
2004-07-19) </dc:creator> <dc:description
xmlns:dc="http://purl.org/dc/elements/1.1/"
xmlns:xsi="http://www.w3.org/2001/XMLSchema-instance"
xsi:schemaLocation="http://purl.org/dc/elements/1.1/
http://dublincore.org/schemas/xmls/simpledc20021212.xsd"> Jhove (Rel. 1.0 (beta 2),
2004-07-19) Date: 2004-12-26 22:08:25 EST RepresentationInformation:
websites/chnm/september11/REPOSITORY/CONTRIBUTORS/steve_levine/WTC009.jpg
ReportingModule: JPEG-hul, Rel. 1.0 (2004-06-23) LastModified: 2004-01-27 13:05:25 EST
Size: 729814 Format: JPEG Version: 1.02 Status: Well-formed and valid MIMEtype:
image/jpeg Profile: JFIF JPEGMetadata: CompressionType: Huffman coding, Baseline DCT
Images: Image: NisoImageMetadata: MIMEType: image/jpeg ByteOrder: big-endian
CompressionScheme: JPEG ColorSpace: YCbCr SamplingFrequencyUnit: inch ImageWidth:
1024 ImageLength: 841 BitsPerSample: 8 SamplesPerPixel: 3 RestartInterval: 128 Scans:
1 QuantizationTables: QuantizationTable: Precision: 8-bit DestinationIdentifier: 0
ApplicationSegments: APP0, APP13, APP14 Checksum: c176826e Type: CRC32 Checksum:
6a18d2726cd6340b076a02cd8738d63 Type: MD5 Checksum:
d3d99e9f5162169c295370cede6fbbc0c4f5b2d2 Type:

```
SHA-1</dc:description></daap:rawOutput>
  </daap:PDI>
  </didl:Statement>
  </didl:Descriptor>
  <didl:Resource mimeType="text/plain; charset=US-ASCII" ref="test/error.xml" />
  </didl:Component>
  </didl:Item>
- <didl:Item>
- <didl:Component>
- <didl:Descriptor>
- <didl:Statement mimeType="text/xml; charset=UTF-8">
- <daap:PDI>
  <daap:signature>file-4.07 magic file from /usr/share/file/magic</daap:signature>
- <daap:context>
  <daap:type>mimeType</daap:type>
  <daap:value>application/msword</daap:value>
  </daap:context>
- <daap:reference>
  <daap:type>identifier</daap:type>
  <daap:value>100.700/929085B8B881699362B8F33D56E2701F</daap:value>
  </daap:reference>
- <daap:reference>
  <daap:type>internalIdentifier</daap:type>
  <daap:value>929085B8B881699362B8F33D56E2701F.0</daap:value>
  </daap:reference>
  <daap:rawOutput>Microsoft Office Document</daap:rawOutput>
  </daap:PDI>
  </didl:Statement>
  </didl:Descriptor>
  <didl:Resource mimeType="text/plain; charset:utf8">file-4.07 magic file from
/usr/share/file/magic</didl:Resource>
  </didl:Component>
- <didl:Component>
- <didl:Descriptor>
- <didl:Statement mimeType="text/xml; charset=UTF-8">
- <daap:PDI>
  <daap:signature>java version "1.4.2_05" Java(TM) 2 Runtime Environment, Standard
Edition (build 1.4.2_05-b04) Java HotSpot(TM) Client VM (build 1.4.2_05-b04, mixed
mode)</daap:signature>
- <daap:reference>
  <daap:type>identifier</daap:type>
  <daap:value>100.700/929085B8B881699362B8F33D56E2701F</daap:value>
  </daap:reference>
- <daap:reference>
  <daap:type>internalIdentifier</daap:type>
  <daap:value>929085B8B881699362B8F33D56E2701F.0</daap:value>
  </daap:reference>
- <daap:fixity>
  <daap:type>SHA</daap:type>
  <daap:value>FC372A0EB00E87AB487B09BC229B8DEB6F7881E0</daap:value>
  </daap:fixity>
- <daap:fixity>
  <daap:type>MD5</daap:type>
  <daap:value>18E054BE01AC75CEA78CC3843630755E</daap:value>
  </daap:fixity>
  <daap:rawOutput>SHA:FC372A0EB00E87AB487B09BC229B8DEB6F7881E0
MD5:18E054BE01AC75CEA78CC3843630755E</daap:rawOutput>
  </daap:PDI>
  </didl:Statement>
  </didl:Descriptor>
  <didl:Resource mimeType="text/plain; charset:utf8">java version "1.4.2_05" Java(TM) 2
Runtime Environment, Standard Edition (build 1.4.2_05-b04) Java HotSpot(TM) Client VM
```

```
(build 1.4.2_05-b04, mixed mode)</didl:Resource>
  </didl:Component>
- <didl:Component>
- <didl:Descriptor>
- <didl:Statement mimeType="text/xml; charset=UTF-8">
- <daap:PDI>
  <daap:signature>Jhove (Rel. 1.0 (beta 2), 2004-07-19) Date: 2005-12-02 11:52:19 EST
App: Module: BYTESTREAM 1.0 OutputHandler: TEXT 1.1 OutputHandler: XML 1.1 Usage:
java Jhove [-c conf] [-m module [-p param]] [-h handler] [-e encoding] [-H handler] [-o
output] [-x saxclass] [-t tempdir] [-b bufsize] [[-krs] file-or-uri] Rights: Copyright
2003-2004 by JSTOR and the President and Fellows of Harvard College. Released under the
GNU Lesser General Public License.</daap:signature>
- <daap:provenance>
  <daap:type>lastModified</daap:type>
  <daap:value>2005-11-27T12:15:53-05:00</daap:value>
  </daap:provenance>
- <daap:context>
  <daap:type>mimeType</daap:type>
  <daap:value>application/octet-stream</daap:value>
  </daap:context>
- <daap:context>
  <daap:type>status</daap:type>
  <daap:value>Well-formed and valid</daap:value>
  </daap:context>
- <daap:context>
  <daap:type>format</daap:type>
  <daap:value>bytestream</daap:value>
  </daap:context>
- <daap:reference>
  <daap:type>identifier</daap:type>
  <daap:value>100.700/929085B8B881699362B8F33D56E2701F</daap:value>
  </daap:reference>
- <daap:reference>
  <daap:type>internalIdentifier</daap:type>
  <daap:value>929085B8B881699362B8F33D56E2701F.0</daap:value>
  </daap:reference>
- <daap:fixity>
  <daap:type>size</daap:type>
  <daap:value>45568</daap:value>
  </daap:fixity>
- <daap:fixity>
  <daap:type>SHA-1</daap:type>
  <daap:value>ce9b772ab35822ac7a8b7d42954b9379fcdde6a6</daap:value>
  </daap:fixity>
- <daap:fixity>
  <daap:type>MD5</daap:type>
  <daap:value>d22cf6a3c8a1573f6d84ca3d19b8872</daap:value>
  </daap:fixity>
- <daap:fixity>
  <daap:type>CRC32</daap:type>
  <daap:value>887b0908</daap:value>
  </daap:fixity>
  <daap:rawOutput><?xml version="1.0" encoding="UTF-8"?> <jhove
xmlns:xsi="http://www.w3.org/2001/XMLSchema-instance"
xmlns="http://hul.harvard.edu/ois/xml/ns/jhove"
xsi:schemaLocation="http://hul.harvard.edu/ois/xml/ns/jhove
http://hul.harvard.edu/ois/xml/xsd/jhove/jhove.xsd" name="Jhove" release="1.0 (beta
2)" date="2004-07-19"> <date>2005-12-02T11:52:19-05:00</date> <repInfo
uri="test/aihtpresentation.ppt"> <reportingModule release="1.0"
date="2004-06-23">BYTESTREAM</reportingModule>
<lastModified>2005-11-27T12:15:53-05:00</lastModified> <size>45568</size>
<format>bytestream</format> <status>Well-formed and valid</status>
<mimeType>application/octet-stream</mimeType> <properties> </properties>
```

<checksums> <checksum type="CRC32">887b0908</checksum> <checksum type="MD5">d22cf6a3c8a1573f6d84ca3d19b8872</checksum> <checksum type="SHA-1">ce9b772ab35822ac7a8b7d42954b9379fcdde6a6</checksum> </checksums> </repInfo> </jhove></daap:rawOutput>
</daap:PDI>
</didl:Statement>
</didl:Descriptor>
<didl:Resource mimeType="text/plain; charset:utf8">Jhove (Rel. 1.0 (beta 2), 2004-07-19) Date: 2005-12-02 11:52:19 EST App: Module: BYTESTREAM 1.0 OutputHandler: TEXT 1.1 OutputHandler: XML 1.1 Usage: java Jhove [-c conf] [-m module [-p param]] [-h handler] [-e encoding] [-H handler] [-o output] [-x saxclass] [-t tempdir] [-b bufsize] [[-krs] file-or-uri] Rights: Copyright 2003-2004 by JSTOR and the President and Fellows of Harvard College. Released under the GNU Lesser General Public License.</didl:Resource>
</didl:Component>
- <didl:Component>
- <didl:Descriptor>
- <didl:Statement mimeType="text/xml; charset=UTF-8">
- <daap:PDI>
<daap:signature>GNU strings 2.15.90.0.3 20040415 Copyright 2004 Free Software Foundation, Inc. This program is free software; you may redistribute it under the terms of the GNU General Public License. This program has absolutely no warranty.</daap:signature>
- <daap:reference>
<daap:type>identifier</daap:type>
<daap:value>100.700/929085B8B881699362B8F33D56E2701F</daap:value>
</daap:reference>
- <daap:reference>
<daap:type>internalIdentifier</daap:type>
<daap:value>929085B8B881699362B8F33D56E2701F.0</daap:value>
</daap:reference>
- <daap:fixity>
<daap:type>rawCharacters</daap:type>
<daap:value>z[ 00 z[ 00 z[ 00 z[ 0 MPEG 21 Framework - Bucket Michael Nelson Johan Bollen Giridhar Manepalli Rabia Haq Digital Item Declaration Digital Item Identification Rights Expressions Language Rights Data Dictionary Bucket Metadata DIDL descriptors Methods DIDL resources Access Control Bucket - Metadata <didl:Descriptor> <didl:Statement mimeType="text/xml; charset=UTF-8"> <dc:creator ">JHOVE</dc:creator> <dc:description "> <?xml version="1.0" encoding="UTF-8" ?> <jhove> <date>2004-10-26T19:30:34-04:00</date> <repInfo uri="email1013.txt"> <reportingModule release="1.0" date="2004-05-05">ASCII-hul</reportingModule> <lastModified>2004-10-26T19:10:09-04:00</lastModified> <mimeType>text/plain; charset=US-ASCII</mimeType> <name>ASCIIMetadata</name> <values arity="List" type="Property"> <checksums> <checksum type="MD5">a59a6af6b1d8bd757282b3a512a0119f</checksum> </checksums> </repInfo> </jhove> </dc:description> </didl:Statement> </didl:Descriptor> Bucket - Methods bucket.xml <didl:Resource mimeType="application/octet-stream" encoding="base64">QgJF87Cgl9CgljbG9zZShDU1MpOwoJcHJpbnQgIi0tPiIsICRicmVhazsK CXByaW50ICI8L3N0WxlPiIsICRicmVhazsKCXByaW50ICI8L2hlYWQ+IiwgJGJyZWFrOwoJcH JpbnQgIjxib2R5PiIs ICRicmVhazsKfQoKc3ViIGVuZF94aHRtbCB7CglwcmludCAiPC9ib2R5PiIsICRicmVhazsKCXBy aW50ICI8L2h0bWw+IjsKfQo= </didl:Resource> REL Elements License is a Grant of Right to Principle on Property based on a Condition Bucket - ACL Access control list Admin rights User rights Staff rights REL to ACL Symbiosis Click to edit Master text styles Second level Third level Fourth level Fifth level Click to edit Master title style Click to edit Master title style Click to edit Master subtitle style Click to edit Master text styles Second level Third level Fourth level Fifth level MPEG 21 gmanepal Watermark gmanepal Microsoft PowerPoint "Arial October 27, 2004 "System "Arial AIHT Conference4 "Arial "Arial MPEG 21 3 "Arial Framework % "Arial "Arial Bucket) "Arial Michael Nelson "Arial Johan "Arial Bollen "Arial Giridhar "Arial Manepallie "Arial Rabiah "Arial z[ 00 z[ 00 z[ 00 z[ 0 MPEG 21 Framework - Bucket Michael Nelson Johan Bollen Giridhar Manepalli Rabia Haq Digital Item Declaration Digital Item Identification Rights Expressions Language Rights Data Dictionary Bucket Metadata DIDL descriptors Methods DIDL resources Access Control Bucket - Metadata <didl:Descriptor> <didl:Statement mimeType="text/xml; charset=UTF-8"> <dc:creator ">JHOVE</dc:creator> <dc:description "> <?xml version="1.0" encoding="UTF-8" ?> <jhove> <date>2004-10-26T19:30:34-04:00</date> <repInfo

uri="email1013.txt"> <reportingModule release="1.0" date="2004-05-05">ASCII-hul</reportingModule> <lastModified>2004-10-26T19:10:09-04:00</lastModified> <mimeType>text/plain; charset=US-ASCII</mimeType> <name>ASCIIMetadata</name> <values arity="List" type="Property"> <checksums> <checksum type="MD5">a59a6af6b1d8bd757282b3a512a0119f</checksum> </checksums> </repInfo> </jhove> </dc:description> </didl:Statement> </didl:Descriptor> Bucket - Methods bucket.xml <didl:Resource mimeType="application/octet-stream" encoding="base64">QgJF87Cgl9CgljbG9zZShDU1MpOwoJcHJpbnQgIi0tPiIsICRicmVhazsKCXByaW50ICI8L3N0WxlPiIsICRicmVhazsKCXByaW50ICI8L2hlYWQ+IiwgJGJyZWFrOwoJcHJpbnQgIjxib2R5PiIs ICRicmVhazsKfQoKc3ViIGVuZF94aHRtbCB7CglwcmludCAiPC9ib2R5PiIsICRicmVhazsKCXByaW50ICI8L2h0bWw+IjsKfQo= </didl:Resource> REL Elements License is a Grant of Right to Principle on Property based on a Condition Bucket - ACL Access control list Admin rights User rights Staff rights REL to ACL Symbiosis z[ 00 z[ 00 z[ 00 z[ 0 MPEG 21 Framework - Bucket Michael Nelson Johan Bollen Giridhar Manepalli Rabia Haq Digital Item Declaration Digital Item Identification Rights Expressions Language Rights Data Dictionary Bucket Metadata DIDL descriptors Methods DIDL resources Access Control Bucket - Metadata <didl:Descriptor> <didl:Statement mimeType="text/xml; charset=UTF-8"> <dc:creator ">JHOVE</dc:creator> <dc:description "> <?xml version="1.0" encoding="UTF-8" ?> <jhove> <date>2004-10-26T19:30:34-04:00</date> <repInfo uri="email1013.txt"> <reportingModule release="1.0" date="2004-05-05">ASCII-hul</reportingModule> <lastModified>2004-10-26T19:10:09-04:00</lastModified> <mimeType>text/plain; charset=US-ASCII</mimeType> <name>ASCIIMetadata</name> <values arity="List" type="Property"> <checksums> <checksum type="MD5">a59a6af6b1d8bd757282b3a512a0119f</checksum> </checksums> </repInfo> </jhove> </dc:description> </didl:Statement> </didl:Descriptor> Bucket - Methods bucket.xml <didl:Resource mimeType="application/octet-stream" encoding="base64">QgJF87Cgl9CgljbG9zZShDU1MpOwoJcHJpbnQgIi0tPiIsICRicmVhazsKCXByaW50ICI8L3N0WxlPiIsICRicmVhazsKCXByaW50ICI8L2hlYWQ+IiwgJGJyZWFrOwoJcHJpbnQgIjxib2R5PiIs ICRicmVhazsKfQoKc3ViIGVuZF94aHRtbCB7CglwcmludCAiPC9ib2R5PiIsICRicmVhazsKCXByaW50ICI8L2h0bWw+IjsKfQo= </didl:Resource> REL Elements License is a Grant of Right to Principle on Property based on a Condition Bucket - ACL Access control list Admin rights User rights Staff rights REL to ACL Symbiosis On-screen Show Arial Times New Roman Wingdings Watermark MPEG 21 Framework - Bucket MPEG 21 MPEG 21 Components Bucket Bucket - Metadata Bucket - Methods REL Elements Bucket - ACL REL to ACL Symbiosis Fonts Used Design Template Slide Titles gmanepal</daap:value>
</daap:fixity>
<daap:rawOutput>z[ 00 z[ 00 z[ 00 z[ 0 MPEG 21 Framework - Bucket Michael Nelson Johan Bollen Giridhar Manepalli Rabia Haq Digital Item Declaration Digital Item Identification Rights Expressions Language Rights Data Dictionary Bucket Metadata DIDL descriptors Methods DIDL resources Access Control Bucket - Metadata <didl:Descriptor> <didl:Statement mimeType="text/xml; charset=UTF-8"> <dc:creator ">JHOVE</dc:creator> <dc:description "> <?xml version="1.0" encoding="UTF-8" ?> <jhove> <date>2004-10-26T19:30:34-04:00</date> <repInfo uri="email1013.txt"> <reportingModule release="1.0" date="2004-05-05">ASCII-hul</reportingModule> <lastModified>2004-10-26T19:10:09-04:00</lastModified> <mimeType>text/plain; charset=US-ASCII</mimeType> <name>ASCIIMetadata</name> <values arity="List" type="Property"> <checksums> <checksum type="MD5">a59a6af6b1d8bd757282b3a512a0119f</checksum> </checksums> </repInfo> </jhove> </dc:description> </didl:Statement> </didl:Descriptor> Bucket - Methods bucket.xml <didl:Resource mimeType="application/octet-stream" encoding="base64">QgJF87Cgl9CgljbG9zZShDU1MpOwoJcHJpbnQgIi0tPiIsICRicmVhazsKCXByaW50ICI8L3N0WxlPiIsICRicmVhazsKCXByaW50ICI8L2hlYWQ+IiwgJGJyZWFrOwoJcHJpbnQgIjxib2R5PiIs ICRicmVhazsKfQoKc3ViIGVuZF94aHRtbCB7CglwcmludCAiPC9ib2R5PiIsICRicmVhazsKCXByaW50ICI8L2h0bWw+IjsKfQo= </didl:Resource> REL Elements License is a Grant of Right to Principle on Property based on a Condition Bucket - ACL Access control list Admin rights User rights Staff rights REL to ACL Symbiosis Click to edit Master text styles Second level Third level Fourth level Fifth level Click to edit Master title style Click to edit Master title style Click to edit Master subtitle style Click to edit Master text styles Second level Third level Fourth level Fifth level MPEG 21 gmanepal Watermark gmanepal Microsoft PowerPoint "Arial October 27, 2004 "System "Arial AIHT Conference4 "Arial "Arial MPEG 21 3 "Arial Framework % "Arial "Arial Bucket) "Arial Michael Nelson "Arial Johan "Arial Bollen "Arial Giridhar "Arial Manepallie "Arial Rabiah "Arial z[ 00 z[ 00 z[ 00 z[ 0 MPEG 21

Framework - Bucket Michael Nelson Johan Bollen Giridhar Manepalli Rabia Haq Digital Item Declaration Digital Item Identification Rights Expressions Language Rights Data Dictionary Bucket Metadata DIDL descriptors Methods DIDL resources Access Control Bucket - Metadata <didl:Descriptor> <didl:Statement mimeType="text/xml; charset=UTF-8"> <dc:creator ">JHOVE</dc:creator> <dc:description "> <?xml version="1.0" encoding="UTF-8" ?> <jhove> <date>2004-10-26T19:30:34-04:00</date> <repInfo uri="email1013.txt"> <reportingModule release="1.0" date="2004-05-05">ASCII-hul</reportingModule> <lastModified>2004-10-26T19:10:09-04:00</lastModified> <mimeType>text/plain; charset=US-ASCII</mimeType> <name>ASCIIMetadata</name> <values arity="List" type="Property"> <checksums> <checksum type="MD5">a59a6af6b1d8bd757282b3a512a0119f</checksum> </checksums> </repInfo> </jhove> </dc:description> </didl:Statement> </didl:Descriptor> Bucket - Methods bucket.xml <didl:Resource mimeType="application/octet-stream" encoding="base64">QgJF87Cgl9CgljbG9zZShDU1MpOwoJcHJpbnQgIi0tPiIsICRicmVhazsKCXBya W50ICI8L3N0WxlPiIsICRicmVhazsKCXByaW50ICI8L2hlYWQ+IiwgJGJyZWFrOwoJcHJpbnQgIjxib2R5PiIs ICRicmVhazsKfQoKc3ViIGVuZF94aHRtbCB7Cglwcml udCAiPC9ib2R5PiIsICRicmVhazsKCXByaW50ICI8L2h0bWw+IjsKfQo= </didl:Resource> REL Elements License is a Grant of Right to Principle on Property based on a Condition Bucket - ACL Access control list Admin rights User rights Staff rights REL to ACL Symbiosis z[ 00 z[ 00 z[ 00 z[ 0 MPEG 21 Framework - Bucket Michael Nelson Johan Bollen Giridhar Manepalli Rabia Haq Digital Item Declaration Digital Item Identification Rights Expressions Language Rights Data Dictionary Bucket Metadata DIDL descriptors Methods DIDL resources Access Control Bucket - Metadata <didl:Descriptor> <didl:Statement mimeType="text/xml; charset=UTF-8"> <dc:creator ">JHOVE</dc:creator> <dc:description "> <?xml version="1.0" encoding="UTF-8" ?> <jhove> <date>2004-10-26T19:30:34-04:00</date> <repInfo uri="email1013.txt"> <reportingModule release="1.0" date="2004-05-05">ASCII-hul</reportingModule> <lastModified>2004-10-26T19:10:09-04:00</lastModified> <mimeType>text/plain; charset=US-ASCII</mimeType> <name>ASCIIMetadata</name> <values arity="List" type="Property"> <checksums> <checksum type="MD5">a59a6af6b1d8bd757282b3a512a0119f</checksum> </checksums> </repInfo> </jhove> </dc:description> </didl:Statement> </didl:Descriptor> Bucket - Methods bucket.xml <didl:Resource mimeType="application/octet-stream" encoding="base64">QgJF87Cgl9CgljbG9zZShDU1MpOwoJcHJpbnQgIi0tPiIsICRicmVhazsKCXBya W50ICI8L3N0WxlPiIsICRicmVhazsKCXByaW50ICI8L2hlYWQ+IiwgJGJyZWFrOwoJcHJpbnQgIjxib2R5PiIs ICRicmVhazsKfQoKc3ViIGVuZF94aHRtbCB7Cglwcml udCAiPC9ib2R5PiIsICRicmVhazsKCXByaW50ICI8L2h0bWw+IjsKfQo= </didl:Resource> REL Elements License is a Grant of Right to Principle on Property based on a Condition Bucket - ACL Access control list Admin rights User rights Staff rights REL to ACL Symbiosis On-screen Show Arial Times New Roman Wingdings Watermark MPEG 21 Framework - Bucket MPEG 21 MPEG 21 Components Bucket Bucket - Metadata Bucket - Methods REL Elements Bucket - ACL REL to ACL Symbiosis Fonts Used Design Template Slide Titles gmanepal</daap:rawOutput>
</daap:PDI>
</didl:Statement>
</didl:Descriptor>
<didl:Resource mimeType="text/plain; charset:utf8">GNU strings 2.15.90.0.3 20040415 Copyright 2004 Free Software Foundation, Inc. This program is free software; you may redistribute it under the terms of the GNU General Public License. This program has absolutely no warranty.</didl:Resource>
<didl:Resource mimeType="application/octet-stream" ref="test/aihtpresentation.ppt" />
</didl:Component>
</didl:Item>
- <didl:Item>
- <didl:Component>
- <didl:Descriptor>
- <didl:Statement mimeType="text/xml; charset=UTF-8">
- <daap:PDI>
<daap:signature>file-4.07 magic file from /usr/share/file/magic</daap:signature>
- <daap:context>
<daap:type>mimeType</daap:type>
<daap:value>text/plain; charset=us-ascii</daap:value>
</daap:context>
- <daap:reference>

```
  <daap:type>identifier</daap:type>
  <daap:value>100.700/5CA400AD12387F6FD23150BA54F2A1B7</daap:value>
  </daap:reference>
- <daap:reference>
  <daap:type>internalIdentifier</daap:type>
  <daap:value>5CA400AD12387F6FD23150BA54F2A1B7.0</daap:value>
  </daap:reference>
  <daap:rawOutput>ASCII English text</daap:rawOutput>
  </daap:PDI>
  </didl:Statement>
  </didl:Descriptor>
  <didl:Resource mimeType="text/plain; charset:utf8">file-4.07 magic file from
/usr/share/file/magic</didl:Resource>
  </didl:Component>
- <didl:Component>
- <didl:Descriptor>
- <didl:Statement mimeType="text/xml; charset=UTF-8">
- <daap:PDI>
  <daap:signature>java version "1.4.2_05" Java(TM) 2 Runtime Environment, Standard
Edition (build 1.4.2_05-b04) Java HotSpot(TM) Client VM (build 1.4.2_05-b04, mixed
mode)</daap:signature>
- <daap:reference>
  <daap:type>identifier</daap:type>
  <daap:value>100.700/5CA400AD12387F6FD23150BA54F2A1B7</daap:value>
  </daap:reference>
- <daap:reference>
  <daap:type>internalIdentifier</daap:type>
  <daap:value>5CA400AD12387F6FD23150BA54F2A1B7.0</daap:value>
  </daap:reference>
- <daap:fixity>
  <daap:type>SHA</daap:type>
  <daap:value>91CAD9D8C2336AF58AE04C2876CB9AD61D0E5510</daap:value>
  </daap:fixity>
- <daap:fixity>
  <daap:type>MD5</daap:type>
  <daap:value>A5E69B225922A3A6CD46A20C94ECF1C1</daap:value>
  </daap:fixity>
  <daap:rawOutput>SHA:91CAD9D8C2336AF58AE04C2876CB9AD61D0E5510
MD5:A5E69B225922A3A6CD46A20C94ECF1C1</daap:rawOutput>
  </daap:PDI>
  </didl:Statement>
  </didl:Descriptor>
  <didl:Resource mimeType="text/plain; charset:utf8">java version "1.4.2_05" Java(TM) 2
Runtime Environment, Standard Edition (build 1.4.2_05-b04) Java HotSpot(TM) Client VM
(build 1.4.2_05-b04, mixed mode)</didl:Resource>
  </didl:Component>
- <didl:Component>
- <didl:Descriptor>
- <didl:Statement mimeType="text/xml; charset=UTF-8">
- <daap:PDI>
  <daap:signature>Jhove (Rel. 1.0 (beta 2), 2004-07-19) Date: 2005-12-02 11:52:20 EST
App: Module: BYTESTREAM 1.0 OutputHandler: TEXT 1.1 OutputHandler: XML 1.1 Usage:
java Jhove [-c conf] [-m module [-p param]] [-h handler] [-e encoding] [-H handler] [-o
output] [-x saxclass] [-t tempdir] [-b bufsize] [[-krs] file-or-uri] Rights: Copyright
2003-2004 by JSTOR and the President and Fellows of Harvard College. Released under the
GNU Lesser General Public License.</daap:signature>
- <daap:provenance>
  <daap:type>lastModified</daap:type>
  <daap:value>2005-12-02T11:50:32-05:00</daap:value>
  </daap:provenance>
- <daap:context>
  <daap:type>mimeType</daap:type>
  <daap:value>text/plain; charset=US-ASCII</daap:value>
```

```
   </daap:context>
- <daap:context>
   <daap:type>status</daap:type>
   <daap:value>Well-formed and valid</daap:value>
   </daap:context>
- <daap:context>
   <daap:type>format</daap:type>
   <daap:value>ASCII</daap:value>
   </daap:context>
- <daap:reference>
   <daap:type>identifier</daap:type>
   <daap:value>100.700/5CA400AD12387F6FD23150BA54F2A1B7</daap:value>
   </daap:reference>
- <daap:reference>
   <daap:type>internalIdentifier</daap:type>
   <daap:value>5CA400AD12387F6FD23150BA54F2A1B7.0</daap:value>
   </daap:reference>
- <daap:fixity>
   <daap:type>size</daap:type>
   <daap:value>250</daap:value>
   </daap:fixity>
- <daap:fixity>
   <daap:type>SHA-1</daap:type>
   <daap:value>91cad9d8c2336af58ae04c2876cb9ad61de5510</daap:value>
   </daap:fixity>
- <daap:fixity>
   <daap:type>MD5</daap:type>
   <daap:value>a5e69b225922a3a6cd46a2c94ecf1c1</daap:value>
   </daap:fixity>
- <daap:fixity>
   <daap:type>CRC32</daap:type>
   <daap:value>d0d525af</daap:value>
   </daap:fixity>
   <daap:rawOutput><?xml version="1.0" encoding="UTF-8"?> <jhove
xmlns:xsi="http://www.w3.org/2001/XMLSchema-instance"
xmlns="http://hul.harvard.edu/ois/xml/ns/jhove"
xsi:schemaLocation="http://hul.harvard.edu/ois/xml/ns/jhove
http://hul.harvard.edu/ois/xml/xsd/jhove/jhove.xsd" name="Jhove" release="1.0 (beta
2)" date="2004-07-19"> <date>2005-12-02T11:52:20-05:00</date> <repInfo
uri="test/test.txt"> <reportingModule release="1.0" date="2004-05-05">ASCII-
hul</reportingModule> <lastModified>2005-12-02T11:50:32-05:00</lastModified>
<size>250</size> <format>ASCII</format> <status>Well-formed and valid</status>
<mimeType>text/plain; charset=US-ASCII</mimeType> <properties> <property>
<name>ASCIIMetadata</name> <values arity="List" type="Property"> <property>
<name>LineEndings</name> <values arity="List" type="String"> <value>LF</value>
</values> </property> </values> </property> </properties> <checksums> <checksum
type="CRC32">d0d525af</checksum> <checksum
type="MD5">a5e69b225922a3a6cd46a2c94ecf1c1</checksum> <checksum
type="SHA-1">91cad9d8c2336af58ae04c2876cb9ad61de5510</checksum>
</checksums> </repInfo> </jhove></daap:rawOutput>
   </daap:PDI>
   </didl:Statement>
   </didl:Descriptor>
   <didl:Resource mimeType="text/plain; charset:utf8">Jhove (Rel. 1.0 (beta 2), 2004-07-19)
Date: 2005-12-02 11:52:20 EST App: Module: BYTESTREAM 1.0 OutputHandler: TEXT 1.1
OutputHandler: XML 1.1 Usage: java Jhove [-c conf] [-m module [-p param]] [-h handler]
[-e encoding] [-H handler] [-o output] [-x saxclass] [-t tempdir] [-b bufsize] [[-krs] file-
or-uri] Rights: Copyright 2003-2004 by JSTOR and the President and Fellows of Harvard
College. Released under the GNU Lesser General Public License.</didl:Resource>
   </didl:Component>
- <didl:Component>
- <didl:Descriptor>
- <didl:Statement mimeType="text/xml; charset=UTF-8">
```

```
- <daap:PDI>
  <daap:signature>GNU strings 2.15.90.0.3 20040415 Copyright 2004 Free Software
Foundation, Inc. This program is free software; you may redistribute it under the terms of
the GNU General Public License. This program has absolutely no
warranty.</daap:signature>
- <daap:reference>
  <daap:type>identifier</daap:type>
  <daap:value>100.700/5CA400AD12387F6FD23150BA54F2A1B7</daap:value>
  </daap:reference>
- <daap:reference>
  <daap:type>internalIdentifier</daap:type>
  <daap:value>5CA400AD12387F6FD23150BA54F2A1B7.0</daap:value>
  </daap:reference>
- <daap:fixity>
  <daap:type>rawCharacters</daap:type>
  <daap:value>am a man of few words, and account the time for action. But, here we go the
little saga for this episode as it unravels. I engaged myself in the profound sciences,
mathematics, to educate myself and to enrich my mind with "thoughts that
matter".</daap:value>
  </daap:fixity>
  <daap:rawOutput>am a man of few words, and account the time for action. But, here we
go the little saga for this episode as it unravels. I engaged myself in the profound sciences,
mathematics, to educate myself and to enrich my mind with "thoughts that
matter".</daap:rawOutput>
  </daap:PDI>
  </didl:Statement>
  </didl:Descriptor>
  <didl:Resource mimeType="text/plain; charset=US-ASCII" ref="test/test.txt" />
  <didl:Resource mimeType="text/plain; charset:utf8">GNU strings 2.15.90.0.3 20040415
Copyright 2004 Free Software Foundation, Inc. This program is free software; you may
redistribute it under the terms of the GNU General Public License. This program has
absolutely no warranty.</didl:Resource>
  </didl:Component>
  </didl:Item>
  </didl:Item>
  </didl:Container>
  </didl:DIDL>
```